\documentclass[aps,pre,eprint,groupedaddress]{revtex4-2}

\usepackage{graphicx}
\usepackage{xcolor}
\usepackage{amsmath,amssymb,amsfonts}

\begin{document}

\title{$Q$-voter model with independence on signed random graphs:\\ homogeneous approximations}

\author{A.\ Krawiecki and T.\ Gradowski}       

\affiliation{Faculty of Physics,
Warsaw University of Technology, \\
Koszykowa 75, PL-00-662 Warsaw, Poland}

\begin{abstract}
The $q$-voter model with independence is generalized to signed random graphs and studied by means of Monte Carlo simulations and theoretically using the mean field approximation and different forms of the pair approximation. In the signed network with quenched disorder, positive and negative signs associated randomly with the links correspond to reinforcing and antagonistic interactions, promoting, respectively, the same or opposite orientations of two-state spins representing agents' opinions; otherwise, the opinions are called mismatched. With probability $1-p$, the agents change their opinions if the opinions of all members of a randomly selected $q$-neighborhood are mismatched, and with probability $p$, they choose an opinion randomly. The model on networks with finite mean degree $\langle k \rangle$ and fixed fraction of the antagonistic interactions $r$ exhibits ferromagnetic transition with varying the independence parameter $p$, which can be first- or second-order, depending on $q$ and $r$, and disappears for large $r$. Besides, numerical evidence is provided for the occurrence of the spin-glass-like transition for large $r$. The order and critical lines for the ferromagnetic transition on the $p$ vs.\ $r$ phase diagram obtained in Monte Carlo simulations are reproduced qualitatively by the mean field approximation. Within the range of applicability of the pair approximation, for the model with $\langle k \rangle$ finite but $\langle k \rangle \gg q$, predictions of the homogeneous pair approximation concerning the ferromagnetic transition show much better quantitative agreement with numerical results for small $r$ but fail for larger $r$. A more advanced signed homogeneous pair approximation is formulated which distinguishes between classes of active links with a given sign connecting nodes occupied by agents with mismatched opinions; for the model with $\langle k \rangle \gg q$ its predictions agree quantitatively with numerical results in a whole range of $r$ where the ferromagnetic transition occurs.
\end{abstract}


\maketitle


\section{Introduction}
\label{sec:intro}

Social opinion formation is a collective phenomenon for which various agent-based models have been proposed, differing by a spectrum of agents' opinions (discrete or continuous), rules governing changes of these opinions, the effect of interactions between agents on their decisions, etc. \cite{Castellano09}. In particular, nonequilibrium models with binary-state dynamics are widely investigated, in which agents are represented by spins with two possible orientations corresponding to opposite opinions. The latter models belong to several paradigmatic classes, e.g., the voter model \cite{Clifford73,Holley75,Sood05,Sood08,Vazquez08,Pugliese09}, the noisy voter model \cite{Granovsky95, Carro16, Peralta18a}, different variants of the noisy $q$-voter model (called also the nonlinear voter model)
\cite{Castellano09a, Nyczka12, Moretti13, Chmiel15, Jedrzejewski17, Peralta18, Jedrzejewski22, Vieira18, Vieira20, Gradowski20, Nowak21}, the majority-vote model  \cite{Oliveira92, Chen15, Chen17, Nowak20, Chen20, Kim21}, the $q$-neighbor Ising model \cite{Jedrzejewski15, Park17, Chmiel17, Chmiel18}, etc. In most cases, interactions considered in these models are reinforcing and promote identical opinions of interacting agents, in analogy with ferromagnetic (FM) interactions in equilibrium spin models. The structure of these interactions is usually described by complex, possibly heterogeneous networks \cite{Carro16, Moretti13, Jedrzejewski17, Peralta18, Jedrzejewski22, Vieira20, Gradowski20, Chen15, Chen17, Nowak20, Chen20, Kim21, Chmiel18} which reflect generic features of empirical networks of social relationships \cite{Albert02, Dorogovtsev08}. In models with a certain degree of stochasticity, e.g., corresponding to human uncertainty in making decisions, the reinforcing interactions between agents lead to the occurrence of a phase transition to an ordered FM-like phase with one dominant opinion as the level of internal noise is varied. Of particular interest are discontinuous transitions, which under certain conditions were observed in most above-mentioned paradigmatic models \cite{Nyczka12, Moretti13, Chmiel15, Jedrzejewski17, Peralta18, Jedrzejewski22, Vieira18, Vieira20, Gradowski20, Nowak21, Chen17, Nowak20, Jedrzejewski15, Park17, Chmiel17, Chmiel18}, characterized by a sudden appearance (disappearance) of the FM phase with decreasing (increasing) the internal noise and by a possible presence of a hysteresis loop. This FM transition can be described by an appropriate mean-field (MF) approximation, but in many cases, better agreement between theoretical predictions and results of Monte Carlo (MC) simulations is achieved by using a more advanced pair approximation (PA) \cite{Vazquez08, Pugliese09, Peralta18a, Jedrzejewski17, Peralta18, Jedrzejewski22, Vieira20, Gradowski20, Chmiel18}.

Recently, some interest has been attracted by models for the opinion formation with a random mixture of the reinforcing and antagonistic interactions between agents \cite{Krawiecki20, Krawiecki21, Baron21, Baron21a}, the latter promoting opposite opinions of the interacting agents, in analogy with antiferromagnetic (AFM) interactions in equilibrium spin models. The underlying network of interactions becomes then a signed network, with positive links corresponding to reinforcing interactions between agents in the connected nodes and negative links corresponding to antagonistic interactions. Hence, the above-mentioned models for the opinion formation with some forms of internal noise become nonequilibrium counterparts for the equilibrium dilute spin glass (SG) models \cite{Sherrington75, Binder86, Mezard87, Nishimori01, Viana85}. In fact, in such models similar phenomena were observed as in the dilute SG models \cite{Viana85}, e.g., the disappearance of the FM transition (in particular, of the discontinuous transition) and the occurrence of an SG-like transition to a phase with short-range local rather than long-range global ordering as the fraction of the antagonistic interactions is increased \cite{Krawiecki20, Krawiecki21}.  

In this paper, a variant of the noisy $q$-voter model called the $q$-voter model with independence \cite{Nyczka12, Chmiel15, Jedrzejewski17, Peralta18, Jedrzejewski22, Gradowski20, Nowak21} on complex signed networks is investigated. In this model, each agent updates his/her opinion according to a modified probabilistic rule which takes into account both the opinions of a subset of $q$ his/her neighbors and the signs of links leading to them; besides, the agent can change opinion randomly (independently), which introduces a certain level of internal noise to the model. This work is a continuation and extension of the recent studies on the related majority-vote \cite{Krawiecki20} and $q$-neighbor Ising models \cite{Krawiecki21} on complex signed networks, and the obtained results are to a large extent similar to those reported in the two latter cases. In particular, in MC simulations, depending on the parameters of the model, discontinuous and continuous FM transitions are observed as the degree of agents' independence is varied, as well as a tricritical point (TCP) separating the transitions of different orders, corresponding to a certain fraction of the antagonistic interactions in the network, and disappearance of the FM transition for a large fraction of the antagonistic interactions. However, the possible SG-like transition in the model under study with a large fraction of the antagonistic interactions is very weak and hard to confirm in simulations. As in Ref.\ \cite{Krawiecki20, Krawiecki21}, the above-mentioned results concerning the FM transition can be to some extent explained using the MF approximation or a simple version of the PA called homogeneous PA (HPA). The former approximation yields quantitatively correct predictions for the model on networks with a large mean degree of nodes, and the latter one also in the case of networks with a moderate (finite, but still substantially larger than $q$) mean degree of nodes and a small to moderate fraction of antagonistic interactions, but its predictions become even qualitatively incorrect for increasing fraction of the antagonistic interactions. In order to resolve the latter discrepancy, in this paper an improved signed HPA (SHPA) is developed, related to the PA for the $q$-voter models with two kinds of agents with different opinion update rules \cite{Jedrzejewski22}. Predictions of the SHPA show much improved quantitative agreement with results of MC simulations of the FM transition in the model under study on networks with a large and moderate mean degree of nodes and for any fraction of the antagonistic interactions. Besides, the SHPA provides heuristic confirmation of the occurrence of the  SG-like transition in the $q$-voter model on signed networks.

The rest of this paper is organized as follows. In Sec.\ \ref{sec:model} the $q$-voter model on complex signed networks is defined, with an appropriately generalized rule for the update of agents' opinions. In Sec.\ \ref{sec:theory} theoretical approaches used in the study of the model are presented in the order of increasing complexity and accuracy of predictions: MFA (Sec.\ \ref{sec:theory_mfa}), HPA (Sec.\ \ref{sec:theory_hpa}) and SHPA (Sec.\ \ref{sec:theory_shpa}). In Sec.\ \ref{sec:results} results of MC simulations of the model under study are presented, numerical evidence for the occurrence of the FM and SG-like transitions is provided, and the observed properties of these transitions are compared with those predicted by the above-mentioned theories (Sec.\ \ref{sec:resFM} and \ref{sec:resSG}). Sec.\ \ref{sec:conclusions} is devoted to summary and conclusions.

\section{The model}

\label{sec:model}

The model considered in this paper is the $q$-voter model with independence on signed random networks, which is an extension of the widely studied stochastic nonlinear $q$-voter model for the social opinion formation \cite{Castellano09a, Nyczka12, Moretti13, Chmiel15, Jedrzejewski17, Peralta18, Jedrzejewski22}. In the model under study, agents are located in nodes of a random network, indexed by $j=1, 2, \ldots N$, with degrees $k_j$, and represented by spins $\sigma_j =\pm 1$ with the orientations up and down corresponding to two opposite opinions on a given subject. The agents interact via links (edges) of the network, and these interactions influence the rate of the opinion changes (spin-flip rate) for each agent, which is a mixture of the demand for unanimity in a randomly selected subset of his/her $q$ neighbors and independence which allows the agent to change opinion without taking into account his/her neighbors' opinions. Decreasing the degree of independence of the agents can lead to the appearance of one dominant opinion, in analogy with the FM transition with decreasing temperature. The extension of the model proposed in this paper consists in the introduction of two kinds of interactions, represented by two kinds of links: reinforcing interactions preferring identical opinions of the interacting agents, represented by positive links in analogy with FM interactions, and antagonistic interactions preferring opposite opinions of the interacting agents, represented by negative links in analogy with AFM interactions. The kind of interaction is randomly associated with each link, and changing the proportion of the two kinds of interactions allows modification of the sort and order of the phase transitions observed in the model; in particular, for a large fraction of antagonistic interactions, the SG-like rather than the FM transition can be expected. In contrast with the related majority-vote \cite{Krawiecki20} and $q$-neighbor Ising \cite{Krawiecki21} models on signed networks, in the $q$-voter model association of the sign with each link has only conventional character and does not directly affect the spin-flip rate which is sensitive to the kind of interaction represented by each link rather than its sign and exact value. Nevertheless, the network of interactions in the model under study is treated as a signed network for convenience. 

In this paper, the $q$-voter model with independence is considered on complex random networks with degree distribution $P(k)$ and mean degree of nodes $\langle k\rangle$. For simplicity, MC simulations are performed and their results are compared with theoretical predictions only for the model on homogeneous and weakly heterogeneous complex networks, e.g.,
random regular graphs (RRGs) with $P(k)=\delta_{k,K}$, $\langle k \rangle = K$, and Erd\"os-R\'enyi graphs (ERGs) with $P(k)={N-1 \choose k}\rho^{k}(1-\rho)^{N-1-k}$, $\langle k\rangle=(N-1)\rho$, $\rho\ll 1$ \cite{Albert02, Erdos59}; however, the derived HPA and SHPA are valid also for more heterogeneous networks, e.g., scale-free networks with $P(k)\propto k^{-\gamma}$, $\gamma >2$ \cite{Albert02, Barabasi99}. With each link in the network, antagonistic or reinforcing interaction is associated with probability $r$ or $1-r$, respectively. Hence, as explained above, the network can be treated as a signed network, with the signs of the links drawn randomly from a two-point distribution with the $-1$ sign occurring with probability $r$ and $+1$ sign with probability $1-r$. For each realization of the random graph, both the structure and signs of the links remain constant during MC simulation (quenched disorder).

In order to define the rule for the opinion change of the agents in the model under study, the concept of mismatched opinions is introduced. The opinions of the two interacting agents are mismatched if they are opposite while the interaction between the agents is reinforcing, or if they are identical while the interaction between the agents is antagonistic. Then, the unanimity rule for the opinion change followed by agents in the $q$-voter model on networks \cite{Castellano09a, Nyczka12, Moretti13, Chmiel15, Jedrzejewski17, Peralta18, Jedrzejewski22}, i.e., that the agent changes opinion (the corresponding spin flips) if opinions of all agents in a randomly selected subset of his/her $q$ neighbors are opposite to his/her opinion, is replaced by a generalized unanimity rule, i.e., that the agent changes opinion if opinions of all agents in a randomly selected subset of his/her $q$ neighbors are mismatched with his/her opinion. Besides, each agent is endowed with a certain degree of independence, such that he/she ignores the above-mentioned generalized unanimity rule and with probability $p$, $0\le p \le1$, called the independence parameter, makes decision randomly \cite{Nyczka12, Moretti13, Chmiel15, Jedrzejewski17, Peralta18}. In the MC simulations, random sequential updating of the spins is performed according to the above-mentioned rules, and a single MC simulation step (MCSS) corresponds to the update of opinions of all $N$ agents, without repetitions. Eventually, in numerical simulations of the $q$-voter model on signed networks, each MCSS is performed as follows.
\begin{itemize}
    \item[(i.)] A node $j$, $1\le j \le N$, with degree $k_j$ is picked randomly.
    \item[(ii.)] A set of its $q$ neighbors ($q$-neighborhood) is chosen randomly and without repetitions. It is assumed that $0<q\le k_j$; otherwise, the node is excluded from the simulation.  
    \item[(iii.)] With probability $1-p$ the picked agent follows a generalized unanimity rule and flips his/her opinion if the opinions of all members of the chosen $q$-neighborhood are mismatched with his/her opinion. 
    \item[(iv.)] With probability $p$, the picked agent behaves independently and flips or preserves his/her opinion with equal probability $p/2$.
    \item[(v.)] Steps (i.)-(iv.) are repeated until all $N$ spins are updated without repetition.
\end{itemize}

\section{Theory}

\label{sec:theory}

\label{sec:theory_mfa}

\subsection{Mean field approximation}

In this section, simple MFA for the FM transition in the $q$-voter model on signed networks is presented, valid for the model on fully connected graphs, and approximately valid for the model on random graphs with a large mean degree of nodes $\langle k\rangle$. It is a straightforward extension of the MFA for the FM transition in the $q$-voter model on a fully connected graph with only reinforcing interactions \cite{Nyczka12, Moretti13}, corresponding to $r=0$ in the distribution of the signs of links. In the MFA, the macroscopic quantity characterizing the model is the concentration $c_{\uparrow}\equiv c$ of spins with orientation up (hence, $c_{\downarrow}= 1-c$), related to the order parameter, the usual magnetization $m$, by $c=(1+m)/2$. It is assumed that the signs of links are not correlated with the orientations of spins in the connected nodes; thus, the probability that a spin with orientation up or down has a neighbor with a mismatched opinion is $(1-r)(1-c)+rc$ or $(1-r)c+r(1-c)$, respectively. Since in a single simulation step a node occupied by spin with orientation up or down is picked with probability $c$ or $1-c$, respectively, and taking into account the rules (ii.-iv.) of Sec.\ \ref{sec:model} determining the dynamics of the model under study, a dynamical equation for the concentration $c$ can be written as a rate equation, 
\begin{eqnarray}
    \frac{\partial c}{\partial t} &=& (1-c)\left\{ (1-p) \left[ (1-r)c +r (1-c)\right]^q +\frac{p}{2}\right\}
    - c\left\{ \left[ (1-r)(1-c)+rc\right]^q +\frac{p}{2}\right\}. 
    \label{eq:cMF}
\end{eqnarray}
The following calculations are slightly simplified by performing a linear change of variables,
\begin{equation}
    \xi = c(1-2r)+r,
\end{equation}
which transforms Eq.\ (\ref{eq:cMF}) into
\begin{equation}
    \frac{\partial \xi}{\partial t} = \left( 1-p\right) \Phi(\xi) +\frac{p}{2}(1-2\xi) \equiv F(\xi)
    \label{eq:xiMF},
\end{equation}
where
\begin{equation}
        \Phi(\xi) \equiv (1-\xi -r) \xi^q 
    - (\xi -r) (1-\xi)^q.
\end{equation}

Fixed points of Eq.\ (\ref{eq:xiMF}) are solutions of the equation $F(\xi)=0$. Different stable fixed points correspond to different, i.e., PM or FM, phases of the model. In particular, for any $p$ a (stable or unstable) fixed point with $\xi=c=1/2$ exists, corresponding to the PM phase. The critical values of the independence parameter and the order of the FM transition for fixed $q$, $r$ and varying $p$ can be obtained by performing the analysis of the stability of the PM fixed point and of the bifurcations of Eq.\ (\ref{eq:xiMF}), as for the HPA in Sec.\ \ref{sec:theory_hpa}. Alternatively \cite{Nyczka12}, equation $F(\xi)=0$ can be solved with respect to $p$, which for fixed $q$, $r$ yields $p$ as a function of $\xi$ only,
\begin{equation}
    p(\xi)= 1+ \frac{1-2\xi}{2\Phi(\xi) -(1-2\xi)}.
\end{equation}
The plot of the inverse function $\xi(p)$ (which cannot be obtained analytically) is rotated with respect to that of $p(\xi)$ by a right angle, thus for any $0\le p\le 1$ the number and positions of the fixed points of Eq.\ (\ref{eq:xiMF}) can be determined from the crossing points of the plot of the function $p(\xi)$ with the line $p={\rm const}$. In particular, at $\xi=c=1/2$ the function $p(\xi)$ is finite and has an extremum, since using de l'Hospitale's rule one, two, and three times, respectively, yields
\begin{eqnarray}
    \lim_{\xi\rightarrow \frac{1}{2}} p(\xi) &=& 
    \frac{q(1-2r)-1}{q(1-2r)-1 +2^{q-1}}, \label{pxi1/2}\\
    \lim_{\xi\rightarrow \frac{1}{2}} 
    \frac{\partial p}{\partial \xi} &=& 0,\\
    \lim_{\xi\rightarrow \frac{1}{2}} 
    \frac{\partial^2 p}{\partial \xi^2} &=&
    \frac{2^{2q-2}}{3}\frac{q(q-1)\left[ (q-2)(1-2r)-3\right]}{\left[ q(1-2r)-1+2^{q-1}\right]^2}.
    \label{d2pdxi2}
\end{eqnarray}
From Eq.\ (\ref{d2pdxi2}) follows that for $q\le 4$ and $r\ge 0$ as well as for $q=5$ and $r>0$ the extremum at $\xi=1/2$ is always a maximum, while for fixed $q\ge 6$ it can be a minimum or a maximum, depending on $r$. Plotting the function $p(\xi)$ reveals that if it has a maximum at $\xi=1/2$, it is a single maximum in the interval $0\le \xi \le1$, and if it has a minimum at $\xi=1/2$, it has also two maxima, which can be evaluated only numerically, one at $0 <\xi <1/2$ and the other one at $1/2 <\xi <1$, symmetric with respect to the minimum.

From the foregoing considerations follows that the MFA for the $q$-voter model with independence on signed networks depending on $q$, $r$ predicts the continuous or discontinuous FM transition with varying the independence parameter $p$. If $q$ and $r$ are such that the function $p(\xi)$ has a maximum at $\xi=c=1/2$, the transition from the PM to the FM phase with decreasing $p$ is second-order and occurs at $p_{c,MFA}^{(FM)}= p(\xi=1/2)$ given by Eq.\ (\ref{pxi1/2}). For $p> p_{c,MFA}^{(FM)}$ the only stable fixed point of Eq.\ (\ref{eq:xiMF}) is the PM one with $\xi=1/2$, and for $p< p_{c,MFA}^{(FM)}$ there are two stable fixed points at $0 <\xi <1/2$ and $1/2 <\xi <1$ (at $0 <c <1/2$ and $1/2 <c <1$, respectively), corresponding to two symmetric FM phases. In particular, for $r=0$
\begin{equation}
    p_{c,MFA}^{(FM)}= \frac{q-1}{q-1+2^{q-1}},
    \label{pcMFAFM0}
\end{equation}
which is the critical value of the independence parameter in the $q$-voter model with independence and purely reinforcing interactions \cite{Nyczka12}. If $q$ and $r$ are such that the function $p(\xi)$ has a minimum at $\xi=c=1/2$, the transition from the PM to the FM phase with decreasing $p$ is first-order. For $p> p_{c2,MFA}^{(FM)}$ (which is the value of the function $p(\xi)$ at the two symmetric maxima) the only stable fixed point of Eq.\ (\ref{eq:xiMF}) is the PM one with $\xi=1/2$; for $p(\xi=1/2) = p_{c1,MFA}^{(FM)}<p< p_{c2,MFA}^{(FM)}$ there are three stable fixed points, separated by two unstable ones: the PM one with $\xi=1/2$ and two symmetric FM ones at $0 <\xi <1/2$ and $1/2 <\xi <1$, so the PM and FM phases coexist and a hysteresis loop is expected to appear as $p$ is varied in opposite directions; finally, for $p< p_{c1,MFA}^{(FM)}$ only the two stable FM fixed points exist, separated by the unstable PM fixed point. 

For fixed $q$ the dependence of the above-mentioned critical values of the independence parameter on the fraction of antagonistic interactions can be plotted on the $p$ vs.\ $r$ phase diagram, which yields the critical lines for the discontinuous FM transition $p_{c1,MFA}^{(FM)}(r)$, $p_{c2,MFA}^{(FM)}(r)$ and for the continuous FM transition $p_{c,MFA}^{(FM)}(r)$. From Eq.\ (\ref{d2pdxi2}) follows that for $q\ge 6$ the transition changes from discontinuous to continuous with increasing $r$ as the second derivative $\left. \frac{\partial^2 p}{\partial \xi^2}\right|_{\xi=1/2}$ changes sign. Thus, the three above-mentioned critical lines meet in a TCP at 
\begin{eqnarray}
    r_{TCP,MFA}^{(FM)} &=&  \frac{q-5}{2(q-2)}, \\
    p_{TCP,MFA}^{(FM)} &=& p_{c,MFA}^{(FM)} (r_{TCP,MFA}^{(FM)}) = \left. p(\xi=1/2)\right|_{r=r_{TCP,MFA}^{(FM)}} = \frac{q+1}{q+1+2^{q-2}(q-2)}.
\end{eqnarray}
Besides, the MFA predicts that a maximum value $r_{max,MFA}^{(FM)}$ of the fraction of repulsive interactions for which the FM transition is possible exists, determined from the condition $p_{c,MFA}^{(FM)}(r_{max,MFA}^{(FM)})=0$ which yields
\begin{equation}
    r_{max,MFA}^{(FM)}= \frac{q-1}{2q}.
    \label{rmaxMF}
\end{equation}

Predictions of the MFA are qualitatively correct for the model on networks with large $\langle k \rangle$ (Sec.\ \ref{sec:resFM}). In particular, the FM transition observed in MC simulations is continuous for $q\le 5$ and $r>0$, while for $q\ge 6$ a TCP separating the discontinuous and continuous FM transition occurs at $r_{TCP,MC}^{(FM)}>0$. There is also a maximum value $r_{max,MC}^{(FM)}$ of the fraction of the antagonistic interactions for which the FM transition appears at $p>0$. However, for any finite $\langle k \rangle$ predictions based on the PA discussed below are quantitatively more correct.

\subsection{Homogeneous pair approximation}

\label{sec:theory_hpa}

\begin{figure}
    \includegraphics[width=0.5\linewidth]{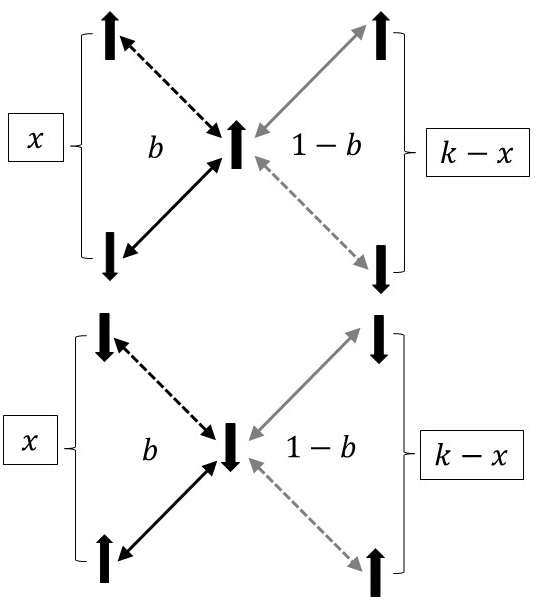}
    \caption{Illustration of the notation used in the derivation of the equations of motion for the macroscopic quantities in the HPA, black arrows denote active links, gray arrows denote inactive links, solid arrows denote reinforcing interactions, dashed arrows denote antagonistic interactions; the central spin with orientation up or down is placed in a node with degree $k$ which has $x$ active and $k-x$ inactive links attached; the concentrations of active and inactive links are $b$ and $1-b$, respectively.}
    \label{fig:links0}
\end{figure}

The PA forms a basis for various more or less detailed theoretical descriptions of the FM transition in models on networks, which can be applied to the $q$-voter model \cite{Jedrzejewski17, Peralta18, Jedrzejewski22, Vieira20, Gradowski20} and are considered to be more accurate than the MFA. A new concept introduced in the PA is that of active links. In the case of models with binary-state dynamics on signed networks, a link is active if it connects nodes occupied by agents with mismatched opinions, as defined in Sec.\ \ref{sec:model}. Thus, active links are reinforcing links connecting nodes occupied by spins with opposite orientations $\nu$, $-\nu$, and antagonistic links connecting nodes occupied by spins with the same orientations $\nu$; the remaining links are called inactive. In other words, the link is active (inactive) if the product of its sign and the signs of the two interacting spins is negative (positive). In this section, the simplest HPA for the $q$-voter models on signed networks is presented, in which all nodes are treated as statistically equivalent, independently of their degrees, and no distinction is made between reinforcing and antagonistic active links (the latter distinction is taken into account in the more advanced SHPA in Sec.\ \ref{sec:theory_shpa}). Eventually, in the framework of the HPA the macroscopic variables describing the $q$-voter model on networks are the concentration $c_{\uparrow}=c$ of nodes occupied by spins with orientation up (normalized to the number of nodes $N$, thus the concentration of nodes occupied by spins with orientation down is $c_{\downarrow}=1-c$) and concentration of active links $b$ (normalized to the total number of links $N\langle k\rangle/2$). Below, a simplified derivation of dynamical equations for the variables $c$, $b$ in the HPA for binary-state models on signed networks is presented (for the notation see Fig.\ \ref{fig:links0}); more details can be found in Ref.\ \cite{Krawiecki20, Krawiecki21}.

A basic assumption in the PA is that orientations of different spins in the neighborhood of a given node are not mutually correlated. Thus, the number of active links $x$ attached to the node with degree $k$ ($x\le k$) occupied by spin with orientation $\nu \in \left\{ \uparrow, \downarrow \right\}$, obeys a binomial distribution $B_{k,x}(\theta_{\nu}) = {k \choose x} \theta_{\nu}^{x} (1-\theta_{\nu})^{k-x}$. Here, $\theta_{\nu}$ is
conditional probability that a link is active provided that it is attached to a randomly chosen node occupied by spin with 
orientation $\nu$. Due to the above-mentioned homogeneous approximation, these probabilities can be expressed in terms of the
macroscopic concentrations $c$, $b$ \cite{Krawiecki20, Krawiecki21},
    \begin{eqnarray}
\theta_{\downarrow} &=& \frac{b-r}{2(1-c)}+r, \label{thetas1}\\
\theta_{\uparrow} &=& \frac{b-r}{2c}+r. \label{thetas2}
\end{eqnarray}

In the framework of the HPA, the dynamical equations for the concentrations $c$, $b$ have a form of rate equations. Let us assume that the flip rate for a spin with any orientation, given that it occupies a node with degree $k$, depends only on the number of active links $x$ attached to this node; this rate is denoted as $f(x|k)$. The average flip rate for spins with orientation $\nu$ is obtained by averaging $f(x|k)$ over the degree distribution of nodes $P(k)$ and over the appropriate distribution $B_{k,x}(\theta_{\nu})$ of the number of active links attached to the node with degree $k$. Thus, 
\begin{eqnarray}
\frac{\partial c}{\partial t} &=& (1-c) \sum_{k} P(k) \sum_{x=0}^{k}B_{k,x} (\theta_{\downarrow}) f(x|k) -
c \sum_{k} P(k)\sum_{x=0}^{k}B_{k,x} (\theta_{\uparrow}) f(x|k).
\label{ratec_HPA}
\end{eqnarray}
If in an elementary simulation step a spin in a node with degree $k$ and $x$ attached active links flips, the $x$ active links become inactive, and $k-x$ links become active, thus the concentration $b$ of active links is changed by
\begin{equation}
\Delta_{b} \left( x\left| k\right.\right)=\frac{2}{N\langle k\rangle}\left( k-2x \right).
\label{deltab_HPA}
\end{equation}   
Since the attempts to flip a spin take place at a rate $1/N$, $N\rightarrow \infty$, the rate equation for the concentration of active links is
\begin{equation}
     \frac{\partial b}{\partial t} = \frac{2}{\langle k\rangle} \sum_{\nu \in \left\{ \uparrow,\downarrow\right\}} c_{\nu}    \sum_{k} P\left( k\right)  
        \sum_{x=0}^{k}
        B_{k,x}\left( \theta_{\nu}\right) 
        f\left( x\left| k\right.\right) \left( k-2x\right).
        \label{rateb_HPA}
\end{equation}

In the case of the $q$-voter model with independence on signed networks, the flip rate for a spin occupying a node with degree $k$ and with attached $x$ active links is
\begin{equation}
    f(x|k) = (1-p) \frac{{x \choose q}}{{k \choose q}} + \frac{p}{2}.
    \label{rate_fxk}
\end{equation}
Inserting Eq.\ (\ref{rate_fxk}) into Eq.\ (\ref{ratec_HPA}) and (\ref{rateb_HPA}) and performing summations as in Ref.\ \cite{Jedrzejewski17} the following system of equations for the macroscopic concentrations $c$, $b$ is obtained,
\begin{eqnarray}
    \frac{\partial c}{\partial t} &=& (1-c) R\left( \theta_{\downarrow}\right) - 
    c R\left( \theta_{\uparrow}\right) \equiv A(c,b),
    \label{dcdt_HPA}\\
     \frac{\partial b}{\partial t} &=& (1-c) \left[ \langle k\rangle R\left( \theta_{\downarrow}\right) - 2S\left( \theta_{\downarrow}\right)\right]
     + c \left[ \langle k\rangle R\left( \theta_{\uparrow}\right) - 2S\left( \theta_{\uparrow}\right)\right] \equiv B(c,b),
     \label{dbdt_HPA}
\end{eqnarray}
where
\begin{eqnarray}
    R (\theta) &=& (1-p)\theta^q +\frac{p}{2},
    \label{funR}\\
    S (\theta) &=& (1-p)\theta^q \left[ (\langle k\rangle-q)\theta +q\right] +\frac{p}{2}\langle k\rangle \theta. \label{funS}
\end{eqnarray}

Fixed points of the system of equations (\ref{dcdt_HPA}, \ref{dbdt_HPA}) are solutions of a system of algebraic equations $A(c,b)=0$, $B(c,b)=0$. Different stable fixed points correspond to different, i.e., PM or FM, phases of the model, and different bifurcations affecting the stability of the fixed points correspond to the discontinuous and continuous FM phase transitions. Predictions of the HPA concerning the FM transition in the $q$-voter model on signed networks are qualitatively similar to those reported previously for the majority-vote model \cite{Krawiecki20} and the $q$-neighbor Ising model \cite{Krawiecki21}, but to a large extent can be  obtained analytically rather than numerically.  

The (stable or unstable) fixed point with $c=1/2$ ($m=0$), which corresponds to the PM phase, for fixed $\langle k\rangle$, $q$ exists in a whole range of the parameters $p$, $0 \le p\le 1$, and $r$, $0\le r\le 1$. 
At this point $\theta_{\downarrow}=\theta_{\uparrow} \equiv \theta =b$ from Eq.\ ( \ref{thetas1}, \ref{thetas2}),
and, as a result, equation $A(c=1/2, b=\theta) =0$ is trivially fulfilled. The value of $\theta$ at the PM fixed point depends on $p$, $r$ and is a solution of the equation $B(c=1/2,\theta)=0$. The stability of the PM fixed point $c=1/2$, $b=\theta$ can be analyzed by evaluating eigenvalues $\lambda_1$, $\lambda_2$ of the Jacobian matrix of the system of equations (\ref{dcdt_HPA}, \ref{dbdt_HPA}) at the fixed point. It can be easily checked that 
$\left. \frac{\partial A}{\partial b}\right|_{c=1/2,b=\theta} = \left. \frac{\partial B}{\partial c}\right|_{c=1/2,b=\theta} =0$, thus the eigenvalues at the PM fixed point are
$\lambda_1 = \left. \frac{\partial A}{\partial c}\right|_{c=1/2,b=\theta}$, $\lambda_2 = \left. \frac{\partial B}{\partial b}\right|_{c=1/2,b=\theta}$. Besides, it can be verified numerically that for any values of the parameters $\langle k \rangle$, $q$, $p$, $r$ there is $\lambda_2 <0$, and $\lambda_1$ can change sign with, e.g., varying $p$ and other parameters fixed. Thus, the critical value $p_c$ at which the PM point becomes unstable, and the corresponding value $b_c=\theta_c$ are obtained as solutions of a system of algebraic equations,
\begin{eqnarray}
\lambda_1 = \left. \frac{\partial A}{\partial c}\right|_{c=1/2,b=\theta}
&=& 2(1-p_c) [q(\theta_c -r)-\theta_c]\theta_{c}^{q-1} -p_c =0, 
\label{lambda1PM_HPA} \\
\left. B(c,b)\right|_{c=1/2,b=\theta} &=& \frac{2}{\langle k \rangle} \left\{ (1-p_c) \theta_{c}^{q} \left[ \langle k\rangle -2q -2 (\langle k\rangle -q)\theta_c\right]  + p_c \frac{\langle k\rangle}{2} (1-2\theta_c)\right\} = 0. 
\label{BPM_HPA}
\end{eqnarray}
In general, there are two solutions for $\theta_c$,
    \begin{equation}
        \theta_{c}^{\pm}= \frac{\langle k\rangle (1+2r) -2 \pm \sqrt{\langle k\rangle^2 (1-2r)^2 -4(\langle k\rangle -1)}}{4(\langle k\rangle -1)},
        \label{thetacpm_HPA}
    \end{equation}
    and two corresponding critical values of $p$,
    \begin{equation}
        p_{c}^{\pm}= \frac{2[q(\theta_c -r)-\theta_c]\theta_{c}^{q-1}}{1+ 2[q(\theta_c -r)-\theta_c]\theta_{c}^{q-1}} \equiv p_{c}^{\pm} (r),
        \label{pcpm_HPA}
    \end{equation}
where the notation $p_{c}^{\pm} (r)$ means that $p_{c}^{\pm}$ are considered as functions of $r$ only, with the parameters $\langle k\rangle$, $q$ fixed. In the interval $0\le r \le 1$ the function $p_{c}^{+} (r)$ is positive and decreasing, while the equation $p_{c}^{-}(r)=0$ has two roots at
\begin{equation}
    r_1 =0, \;\;\;\; r_2 = \frac{(q-1)(2q -\langle k \rangle)}{2q(q-\langle k \rangle)}. 
\end{equation}
Thus, for $q<\langle k\rangle/2$ the function $p_{c}^{-}(r)$ for $0<r<r_2$ is negative and for $r> r_2$ is positive and increasing function of $r$; while for $\langle k\rangle/2< q< \langle k\rangle$ the function $p_{c}^{-}(r)$ for $r>0$ is positive and increasing. The two solutions $p_{c}^{\pm}(r)$ merge at $r=r^{\star}$ which from Eq.\ (\ref{thetacpm_HPA}) is
\begin{equation}
    r^{\star} = \frac{1}{2}\left( 1- \frac{2\sqrt{\langle k\rangle -1}}{\langle k \rangle}\right).
    \label{rstar_HPA}
\end{equation}
Hence, the HPA predicts that in the range of parameters $0\le r \le 1$, $0\le p\le 1$ the PM fixed point with $c=1/2$ $(m=0)$, $b=\theta$ is stable for $p>p^{+}(r)$ if $r\le r^{\star}$ and for any $p$ if $r> r^{\star}$; besides it is stable for $p<p^{-}(r)$ if $q\le \langle k \rangle/2$ and $r_2 < r\le r^{\star}$ as well as if $q> \langle k \rangle/2$ and $0\le r\le r^{\star}$.

For $\langle k\rangle \rightarrow \infty$ predictions concerning the transition from the PM to the FM phase with fixed $r$ and decreasing $p$ obtained from the HPA and MFA coincide for any $q$. For the model on signed networks with finite $\langle k \rangle$, provided that $q\ll \langle k\rangle$ predictions of the HPA and MFA are still qualitatively similar, with quantitative differences becoming more pronounced with decreasing $\langle k\rangle$ or increasing $q$. In particular, for $r<r^{\star}$ the FM transition can be second- or first-order, depending on the parameters $\langle k\rangle$, $q$, $r$. In the case of the second-order transition the PM fixed point loses stability via a supercritical pitchfork bifurcation at $p=p_{c,HPA}^{(FM)}=p_{c}^{+}$ and for $p< p_{c,HPA}^{(FM)}$ a pair of stable equilibria with $c >1/2$ ($m>0$), $b<1/2$, or $c<1/2$ ($m<0$), $b<1/2$ emerges, corresponding to the FM phase with positive or negative magnetization, respectively. In the case of the first-order transition as $p$ is decreased two pairs of stable and unstable equilibria appear via two saddle-node bifurcations taking place simultaneously at $p=p_{c2,HPA}^{(FM)}>p_{c}^{+}$,
which can be determined only numerically. For $p_{c}^{+} < p< p_{c2,HPA}^{(FM)}$ the two above-mentioned stable equilibria, one with $c >1/2$ ($m>0$), $b<1/2$, and the other with $c<1/2$ ($m<0$), $b<1/2$, corresponding again to the FM phase with positive or negative magnetization, respectively, coexist with the stable equilibrium with $c=1/2$ ($m=0$), $b \le 1/2$ corresponding to the PM phase;
the basins of attraction of the three stable equilibria are separated by stable manifolds of the two unstable equilibria. Eventually at $p=p_{c1,HPA}^{(FM)}=p_{c}^{+}$ the fixed point corresponding to the PM phase loses stability via a subcritical pitchfork bifurcation by colliding with the above-mentioned pair of unstable equilibria, and for $p<p_{c1,HPA}^{(FM)}$ the only two stable fixed points are those corresponding to the FM phase. Hence, 
for $p_{c1,HPA}^{(FM)}< p< p_{c2,HPA}^{(FM)}$ stable PM and FM phases coexist and a hysteresis loop is expected to appear as $p$ is varied in opposite directions. Possibly, for given $q$ the critical lines $p_{c,HPA}^{(FM)}(r)$ corresponding to the second-order FM transition and  $p_{c1,HPA}^{(FM)}(r)$, $p_{c2,HPA}^{(FM)}(r)$ corresponding to the first-order FM transition meet in a TCP $\left( r_{TCP,HPA}^{(FM)}, p_{TCP,HPA}^{(FM)}\right)$ separating regions in which the FM phase emerges in different ways. The location of this TCP can be determined only numerically. For finite $\langle k\rangle$ and small to moderate $r$ the above-mentioned predictions of the HPA show better quantitative agreement with results of MC simulations of the FM transition in the model under study than those of the MFA (Sec.\ \ref{sec:resFM}), but become incorrect for $r\rightarrow r^{\star}$.

If $q$ and $\langle k\rangle$ are comparable predictions concerning the FM transition  with decreasing $p$ based on the HPA are qualitatively different from those based on the MFA. For $q > \langle k\rangle /2$ from the HPA follows that the FM transition is always second-order and occurs for $0< r< r^{\star}$ at $p=p_{c,HPA}^{(FM)}=p_{c}^{+}$. Unexpectedly, it can be followed by another first-order transition with a hysteresis loop of non-zero width, leading to a sudden increase of $|m|$. These predictions differ significantly from results of MC simulations in the whole range of $r$ (Sec.\ \ref{sec:resFM}).

Another prediction of the HPA, which has no analogy in the MFA, is that for $r_2< r< r^{\star}$ (if $q< \langle k\rangle/2$) or for $0< r< r^{\star}$ (if $q> \langle k\rangle/2$) as $p$ is further reduced the two symmetric stable fixed points with $m>0$ or $m<0$ and $b<1/2$ which exist for $p<p_{c,HPA}^{(FM)}$ or $p<p_{c2,HPA}^{(FM)}$, corresponding to the FM phase, approach each other and eventually at $p=p_{c,HPA}^{\prime (FM)} =p_{c}^{-}$ the PM fixed point with $m=0$ regains stability via inverse supercritical pitchfork bifurcation. This means that for a range of $r$ below $r^{\star}$ for given $q$ the HPA predicts the occurrence of another critical line $p_{c,HPA}^{\prime (FM)}(r)$ corresponding to a continuous transition from the FM to the PM phase with decreasing $p$. This line merges with that for the usual continuous transition from the PM to the FM phase $p_{c,HPA}^{(FM)}(r)$ at $r=r^{\star}$, and the two critical lines form a characteristic cusp marking the border of stability of the FM phase on the $(r,p)$ phase diagram. The occurrence of the transition from the FM to the PM phase for $p\rightarrow 0$ is not confirmed by MC simulations (Sec.\ \ref{sec:resFM}); moreover, this transition is not predicted by a more exact SHPA (Sec.\ \ref{sec:theory_shpa}) which shows better agreement with results of MC simulations for larger $r$. In the case of the majority-vote model \cite{Krawiecki20} and the $q$-neighbor Ising model \cite{Krawiecki21} on signed networks it was speculated that the presence of the additional critical line similar to $p_{c,HPA}^{\prime (FM)} (r)$ is related to the possibility of destabilization of the FM phase and occurrence of the SG-like phase. However, in view of the results of the SHPA it is rather the outcome of the crude approximation made in the HPA which does not distinguish between concentrations of the reinforcing and antagonistic active links, which are characterized by a single concentration $b$.

\subsection{Signed homogeneous pair approximation}

\label{sec:theory_shpa}

\begin{figure}
    \includegraphics[width=0.5\linewidth]{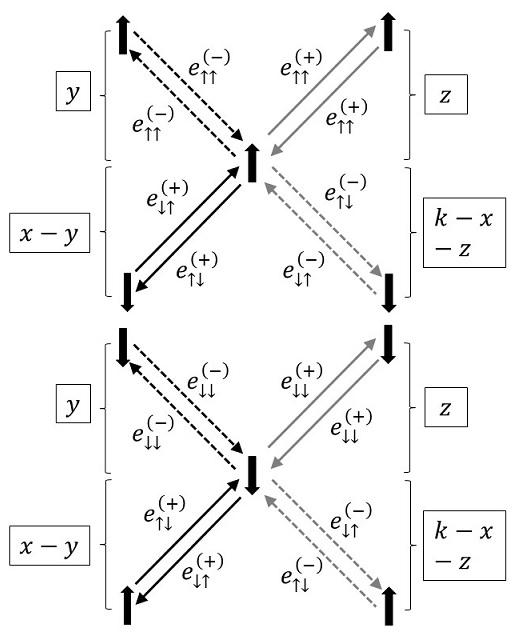}
    \caption{Illustration of the notation used in the derivation of the equations of motion for the macroscopic quantities in the SHPA, black arrows denote active directed links, gray arrows denote inactive directed links, solid arrows denote reinforcing interactions, dashed arrows denote antagonistic interactions associated with directed links, pairs of directed links with opposite directions form links; the central spin with orientation up or down is placed in a node with degree $k$ which has $x$ active and $k-x$ inactive directed links attached, among $x$ active directed links there are $y$ antagonistic directed links (their concentrations are $e_{\uparrow,\uparrow}^{(-)}$ and $e_{\downarrow,\downarrow}^{(-)}$) and $x-y$ reinforcing directed links (their concentrations are $e_{\uparrow,\downarrow}^{(+)}$ and $e_{\downarrow,\uparrow}^{(+)}$), among $k-x$ inactive directed links there are $z$ reinforcing directed links (their concentrations are $e_{\uparrow,\uparrow}^{(+)}$ and $e_{\downarrow,\downarrow}^{(+)}$) and $k-x-z$ antagonistic directed links (their concentrations are $e_{\uparrow,\downarrow}^{(-)}$ and $e_{\downarrow,\uparrow}^{(-)}$).}
    \label{fig:links}
\end{figure}

The signed homogeneous PA (SHPA) derived below generalizes the HPA of Sec.\ \ref{sec:theory_hpa} in such a way that the nodes of the network are still treated as statistically equivalent, but links are divided into classes consisting of links with a given sign connecting spins with given orientations. Thus, in the SHPA the dynamical variables are the concentration of nodes occupied by spins with orientation up and concentrations of active and inactive links corresponding to reinforcing or antagonistic interactions between spins with given orientations at their ends. The rate equations for the dynamical variables can be obtained in a similar way as in the case of the PA for the $q$-voter model with quenched disorder on networks \cite{Jedrzejewski22}; it should be emphasized that in the latter case, the quenched disorder originated from random assignment of two different opinion update rules to the agents rather than opposite signs to the links, which leads to a different formulation of the PA. 

Henceforth it is convenient to consider each link connecting nodes with spins with orientations $\nu, \nu' \in \left\{ \downarrow, \uparrow\right\}$ as a junction of two directed links, one attached to the node occupied by the spin with orientation $\nu$ and pointing at the node occupied by the spin with orientation $\nu'$, and the other one in the opposite direction. The directed links can be reinforcing (positive) or antagonistic (negative), according to the kind (sign) of the link they belong to; thus, junctions of directed links with the same signs are only allowed, forming reinforcing (positive) or antagonistic (negative) links, and junctions of directed links with opposite signs are forbidden. The concentrations of reinforcing and antagonistic directed links attached to the nodes with the spins with orientation $\nu$ and pointing at the nodes with spins with orientation $\nu'$ are denoted as $e_{\nu,\nu'}^{(+)}$ and  $e_{\nu,\nu'}^{(-)}$, respectively; obviously, $e_{\nu,\nu'}^{(+)} = e_{\nu',\nu}^{(+)}$ and $e_{\nu,\nu'}^{(-)} = e_{\nu',\nu}^{(-)}$. Since the concentration of reinforcing directed links (with respect to the total number of directed links in the network $2N\langle k\rangle$) is $1-r$, and that of antagonistic directed links is $r$, there is
\begin{eqnarray}
    e^{(+)}_{\uparrow,\uparrow}+  e^{(+)}_{\downarrow,\downarrow} +
     e^{(+)}_{\uparrow,\downarrow} +  e^{(+)}_{\downarrow,\uparrow} &=& e^{(+)}_{\uparrow,\uparrow}+  e^{(+)}_{\downarrow,\downarrow} +
     2 e^{(+)}_{\uparrow,\downarrow} =1-r,
     \label{sumep}\\
    e^{(-)}_{\uparrow,\uparrow}+  e^{(-)}_{\downarrow,\downarrow} +
     e^{(-)}_{\uparrow,\downarrow} +  e^{(-)}_{\downarrow,\uparrow} &=& e^{(-)}_{\uparrow,\uparrow}+  e^{(-)}_{\downarrow,\downarrow} +
     2 e^{(-)}_{\uparrow,\downarrow} = r.
     \label{sumem}
\end{eqnarray}
Thus, the number of dynamical variables can be reduced by expressing, e.g., concentrations of directed links connecting nodes with spins with opposite orientations by those of directed links connecting nodes with spins with the same orientations,
\begin{eqnarray}
    e^{(+)}_{\uparrow,\downarrow} &=& \frac{1}{2} \left( 1-r- e^{(+)}_{\uparrow,\uparrow}-   e^{(+)}_{\downarrow,\downarrow}\right),
    \label{epud}\\
    e^{(-)}_{\uparrow,\downarrow} &=& \frac{1}{2} \left( r- e^{(-)}_{\uparrow,\uparrow}-   e^{(-)}_{\downarrow,\downarrow}\right).
    \label{emud}
\end{eqnarray}
Besides, the total concentration of directed links attached to the nodes occupied by spins with orientation $\nu$ is equal to the concentration of nodes occupied by spins with orientation $\nu$,
\begin{equation}
  \sum_{\nu' \in \left\{ \downarrow, \uparrow \right\}} \left( e_{\nu,\nu'}^{(+)}+ e_{\nu,\nu'}^{(-)}\right) =c_{\nu}, \;\; \nu \in \left\{\downarrow, \uparrow\right\}.
  \label{sumc}
\end{equation}

In the SHPA again distinction is made between active and inactive links treated as junctions of, respectively, two active and two inactive directed links with opposite directions. Hence, active (inactive) directed links attached to a node occupied by an agent with a given opinion point at nodes occupied by agents with mismatched (matched) opinions. As in the HPA it is assumed that the number $x$ of active directed links attached to a node with degree $k$ occupied by spin with orientation $\nu$ obeys binomial distribution $B_{k,x}(\alpha_{\nu})$. The probabilities $\alpha_{\nu}$, analogous to the probabilities $\theta_{\nu}$ given by Eq.\ (\ref{thetas1}, \ref{thetas2}) in the HPA, can be expressed by the concentrations of directed links, 
\begin{eqnarray}
    \alpha_{\downarrow} &=& \frac{e^{(+)}_{\downarrow, \uparrow} +e^{(-)}_{\downarrow,\downarrow} }{e^{(+)}_{\downarrow, \uparrow}+e^{(-)}_{\downarrow, \uparrow}+e^{(+)}_{\downarrow, \downarrow} + e^{(-)}_{\downarrow, \downarrow}} = \frac{e^{(+)}_{\uparrow, \downarrow} +e^{(-)}_{\downarrow,\downarrow}}{1-c},\label{alphas1} \\
    \alpha_{\uparrow} &=& \frac{e^{(+)}_{\uparrow, \downarrow} +e^{(-)}_{\uparrow,\uparrow} }{e^{(+)}_{\uparrow, \downarrow}+e^{(-)}_{\uparrow, \downarrow}+e^{(+)}_{\uparrow, \uparrow} + e^{(-)}_{\uparrow, \uparrow}} = \frac{e^{(+)}_{\uparrow, \downarrow} +e^{(-)}_{\uparrow,\uparrow}}{c}, \label{alphas2}
\end{eqnarray}
where again $c=c_{\uparrow}$, $1-c = c_{\downarrow}$, $e^{(+)}_{\uparrow, \downarrow}$ is given by Eq.\ (\ref{epud}) and Eq.\ (\ref{sumc}) was used.

In contrast with the HPA, in the SHPA further distinction is made between reinforcing and antagonistic active as well as inactive directed links. Let us consider a node with degree $k$ occupied by spin with orientation $\nu$ and with $x$ active directed links attached. It is assumed that the probability that among these $x$ active links there are $y$ antagonistic links leading to nodes occupied by spins with the same orientation $\nu$ (thus, the remaining $x-y$ active links are reinforcing and lead to nodes occupied by spins with the opposite orientation $-\nu$) is given by the binomial distribution $B_{x,y}\left( \beta_{\nu,\nu}\right)$; and, similarly, that the probability that among the $k-x$ inactive links there are $z$ reinforcing links leading to nodes occupied by spins with the same orientation $\nu$  (thus, the remaining $k-x-z$ inactive links are antagonistic and lead to nodes occupied by spins with opposite orientation $-\nu$) is given by the binomial distribution $B_{k-x,z}\left( \gamma_{\nu,\nu}\right)$; this notation is summarized in Fig.\ \ref{fig:links}. The conditional probabilities $\beta_{\nu,\nu}$ and $\gamma_{\nu,\nu}$ can be expressed by the concentrations of directed links,
\begin{eqnarray}
    \beta_{\uparrow,\uparrow} &=& \frac{e^{(-)}_{\uparrow,\uparrow}}{e^{(+)}_{\uparrow,\downarrow}+ e^{(-)}_{\uparrow,\uparrow}},\\
    \beta_{\downarrow,\downarrow} &=& \frac{e^{(-)}_{\downarrow,\downarrow}}{e^{(+)}_{\downarrow,\uparrow}+ e^{(-)}_{\downarrow,\downarrow}} = \frac{e^{(-)}_{\downarrow,\downarrow}}{e^{(+)}_{\uparrow,\downarrow}+ e^{(-)}_{\downarrow,\downarrow}},\\
    \gamma_{\uparrow,\uparrow} &=& \frac{e^{(+)}_{\uparrow,\uparrow}}{e^{(+)}_{\uparrow,\uparrow}+ e^{(-)}_{\uparrow,\downarrow}},\\
    \gamma_{\downarrow,\downarrow} &=& \frac{e^{(+)}_{\downarrow,\downarrow}}{e^{(-)}_{\downarrow,\uparrow}+ e^{(+)}_{\downarrow,\downarrow}} = 
   \frac{e^{(+)}_{\downarrow,\downarrow}}{e^{(-)}_{\uparrow,\downarrow}+ e^{(+)}_{\downarrow,\downarrow}},
\end{eqnarray}
where $e^{(+)}_{\uparrow, \downarrow}$, $e^{(-)}_{\uparrow, \downarrow}$ are given by Eq.\ (\ref{epud}) and Eq.\ (\ref{emud}), respectively.

In the framework of the SHPA the dynamical equations for the concentration of nodes occupied by spins with orientation up $c$ and concentrations of directed links belonging to different classes again have the form of the rate equations. The general rate equation for $c$ is obviously (\ref{ratec_HPA}) with the conditional probabilities $\theta_{\nu}$ (\ref{thetas1}, \ref{thetas2}) replaced with the corresponding $\alpha_{\nu}$ (\ref{alphas1}, \ref{alphas2}), which for the model under study results in
\begin{equation}
        \frac{\partial c}{\partial t} = (1-c) R\left( \alpha_{\downarrow}\right) - 
    c R\left( \alpha_{\uparrow}\right),
    \label{dcdt_SHPA}\\
\end{equation}
In order to obtain the rate equation for, e.g., the concentration $e^{(+)}_{\uparrow,\uparrow}$, let us consider the average change of this concentration in a single simulation step. For this purpose, two cases must be considered. In the first case, a node occupied by a spin with orientation up is selected, with degree $k$ and with $x$ active and $k-x$ inactive directed links attached, of which, respectively, $y$ are antagonistic and $z$ are reinforcing directed links, which happens with probability $c P(k) B_{k,x}(\alpha_{\uparrow}) B_{x,y}\left( \beta_{\uparrow,\uparrow}\right) B_{k-x,z}\left( \gamma_{\uparrow,\uparrow}\right)$; this spin-flips with probability $f(x|k)$; due to this flip $2z$ reinforcing inactive links between nodes occupied by spins with orientation up are turned into reinforcing active links between nodes occupied by spins with orientation up and down (Fig.\ \ref{fig:links}) which decreases the concentration $e^{(+)}_{\uparrow,\uparrow}$ by $\Delta_1 e^{(+)}_{\uparrow,\uparrow} = - 2z/(N\langle k \rangle)$. In the second case, a node with the above-mentioned properties is selected, occupied by a spin with orientation down, which happens with probability $(1-c) P(k) B_{k,x}(\alpha_{\downarrow}) B_{x,y}\left( \beta_{\downarrow,\downarrow}\right) B_{k-x,z}\left( \gamma_{\downarrow,\downarrow}\right)$; this spin-flips with probability $f(x|k)$; due to this flip $2(x-y)$ reinforcing active links between nodes occupied by spins with orientation up and down are turned into reinforcing inactive links between nodes occupied by spins with orientation up (Fig.\ \ref{fig:links})  which increases the concentration $e^{(+)}_{\uparrow,\uparrow}$ by $\Delta_2 e^{(+)}_{\uparrow,\uparrow} = 2(x-y)/(N\langle k \rangle)$. Taking into account that the attempts to flip a spin take place at a rate $\Delta t = 1/N$, $N \rightarrow \infty$, and averaging over the above-mentioned appropriate probability distributions the following rate equation for the concentration $e^{(+)}_{\uparrow,\uparrow}$ is obtained,
\begin{eqnarray}
    \frac{d e^{(+)}_{\uparrow,\uparrow}}{dt} &=&
    \frac{\Delta_1 e^{(+)}_{\uparrow,\uparrow} + \Delta_2 e^{(+)}_{\uparrow,\uparrow} }{\Delta t} \nonumber\\
    &=& -\frac{2c}{\langle k\rangle} \sum_{k} P(k) 
    \sum_{x=0}^{k} B_{k,x}(\alpha_{\uparrow}) 
    \sum_{y=0}^{x} B_{x,y}\left( \beta_{\uparrow,\uparrow}\right)
    \sum_{z=0}^{k-x} B_{k-x,z}\left( \gamma_{\uparrow,\uparrow}\right) f(x|k) z \nonumber\\
    && + \frac{2(1-c)}{\langle k \rangle} \sum_{k} P(k)  \sum_{x=0}^{k} B_{k,x}(\alpha_{\downarrow})  
    \sum_{y=0}^{x} B_{x,y}\left( \beta_{\downarrow,\downarrow}\right) 
    \sum_{z=0}^{k-x} B_{k-x,z}\left( \gamma_{\downarrow,\downarrow}\right) f(x|k) (x-y).
\end{eqnarray}
Performing summations as in Sec.\ \ref{sec:theory_hpa}, with $f(x|k)$ given by Eq.\ (\ref{rate_fxk}), and considering in a similar way changes of the remaining significant concentrations of directed links in a single simulation step, the following system of equations for the latter quantities is obtained,
\begin{eqnarray}
    \frac{d e^{(+)}_{\uparrow,\uparrow}}{dt} &=&
     -\frac{2c}{\langle k\rangle} \gamma_{\uparrow,\uparrow} \left[ \langle k\rangle R \left( \alpha_{\uparrow} \right) - S \left( \alpha_{\uparrow} \right)\right] +
     \frac{2(1-c)}{\langle k \rangle}
     \left(1- \beta_{\downarrow,\downarrow} \right) 
     S \left( \alpha_{\downarrow} \right),
     \label{depuudt_SHPA}\\
\frac{d e^{(+)}_{\downarrow,\downarrow}}{dt} &=&
\frac{2c}{\langle k\rangle} 
\left(1- \beta_{\uparrow,\uparrow} \right) 
     S \left( \alpha_{\uparrow} \right)
     - \frac{2(1-c)}{\langle k \rangle} \gamma_{\downarrow,\downarrow} 
     \left[ \langle k\rangle R \left( \alpha_{\downarrow} \right) - S \left( \alpha_{\downarrow} \right)\right] 
     \label{depdddt_SHPA}\\
 \frac{d e^{(-)}_{\uparrow,\uparrow}}{dt} &=&    
 - \frac{2c}{\langle k\rangle} \beta_{\uparrow,\uparrow} 
 S \left( \alpha_{\uparrow} \right) +
 \frac{2(1-c)}{\langle k \rangle}
 \left(1- \gamma_{\downarrow,\downarrow} \right) \left[ \langle k\rangle R \left( \alpha_{\downarrow} \right) - S \left( \alpha_{\downarrow} \right)\right],
 \label{demuudt_SHPA}\\
  \frac{d e^{(-)}_{\downarrow,\downarrow}}{dt} &=& 
 \frac{2c}{\langle k\rangle}
 \left(1- \gamma_{\uparrow,\uparrow} \right) \left[ \langle k\rangle R \left( \alpha_{\uparrow} \right) - S \left( \alpha_{\uparrow} \right)\right]
 -\frac{2(1-c)}{\langle k \rangle} \beta_{\downarrow,\downarrow}
 S \left( \alpha_{\downarrow} \right),
 \label{demdddt_SHPA}
\end{eqnarray}
where the functions $R$, $S$ are given by Eq.\ (\ref{funR}) and (\ref{funS}). Natural initial conditions for the five-dimensional system of equations (\ref{dcdt_SHPA},\ref{depuudt_SHPA}-\ref{demdddt_SHPA}) are $c(0)=c_0$, $e^{(+)}_{\uparrow,\uparrow} (0)= c_{0}^{2} (1-r)$, $e^{(+)}_{\downarrow,\downarrow} (0)= (1-c_0)^2 (1-r)$, $e^{(-)}_{\uparrow,\uparrow} (0)=c_{0}^{2}r$, $e^{(-)}_{\downarrow,\downarrow} (0)= (1-c_0)^2 r$, where $0< c_0 <1$ is arbitrary; with the above-mentioned initial conditions, Eq.\ (\ref{sumc}) is obviously fulfilled. 

Fixed points of the system of equations (\ref{dcdt_SHPA}, \ref{depuudt_SHPA}-\ref{demdddt_SHPA}) and their stability can be determined numerically, either using standard numerical tools or by observing long-time asymptotic values of the magnetization $m$ obtained with various $c_0$. In particular, the (stable or unstable) PM fixed point with $c=1/2$ ($m=0$) exists in the whole range of parameters $0\le r\le 1$, $0<p<1$. For fixed $\langle k \rangle$, $q$ and small to moderate $r$ bifurcations affecting its stability are the same as in Eq.\ (\ref{dcdt_HPA}, \ref{dbdt_HPA}). Thus, predictions of the SHPA concerning the order of the phase transition from the PM to the FM phase with decreasing the independence parameter $p$ as well as the corresponding critical values $p_{c1,SHPA}^{(FM)}$, $p_{c2,SHPA}^{(FM)}$ (for the first-order transition) and $p_{c,SHPA}^{(FM)}$ (for the second-order transition) are also close to those of the HPA. Hence, again for finite $\langle k \rangle$, small to moderate $r$ and $q \ll \langle k \rangle$ predictions of the SHPA show better agreement with results of the MC simulations of the FM transition in the model under study than those obtained in the MFA, while for $q$ comparable with $\langle k \rangle$ they differ significantly from results of MC simulations in the whole range of $r$ (Sec.\ \ref{sec:resFM}). 

The main qualitative difference with the predictions of the HPA is that in the case of the SHPA the additional transition from the FM to the PM phase for small $p\rightarrow 0$ is absent, i,e., for given $q$ there is no additional critical line similar to $p_{c,HPA}^{\prime (FM)} (r)$ and the FM phase remains stable for $0< p< p_{c2,SHPA}^{(FM)}$ or $0< p< p_{c,SHPA}^{(FM)}$ in the case of the first- and second-order FM transition, respectively. Thus, the range of $r$ for which the SHPA predicts the FM transition is not constrained to the interval $0< r <r^{\star}$ and the region of stability of the FM phase on the $(r,p)$ phase diagram is not bounded by the cusp characteristic for the HPA. In contrast, for given $q$ the critical line $p_{c,SHPA}^{(FM)} (r)$ usually extends toward larger values of $r$, and for $q \ll \langle k \rangle$ its course agrees quantitatively with the critical line for the FM transition obtained from the MC simulations (Sec.\ \ref{sec:resFM}). Hence, taking into account differences between concentrations of the reinforcing and antagonistic active and inactive directed links leads to much improved quantitative agreement between theoretical predictions of the SHPA and results of the MC simulations for larger fractions of the antagonistic links $r$.
 
\section{Results and discussion}

\label{sec:results}

\subsection{Details of Monte Carlo simulations and analysis of results}

\label{sec:resMET}

In order to verify the occurrence of the FM or SG-like phase transition, MC simulations of the $q$-voter model
under study were performed, and their results were compared with predictions of the MFA, HPA and SHPA from Sec.\ \ref{sec:theory}. In this section results of simulations of the model on RRGs are only presented; in most cases, results for the model on ERGs with the same parameters $\langle k\rangle$, $q$, $r$ are quantitatively similar. Simulations were performed on networks with the number of nodes $10^{3}\le N\le 10^{4}$
using a simulated annealing algorithm with random sequential updating of the agents' opinions, as described in Sec.\ \ref{sec:model}. For each realization of the network and attribution of the reinforcing and antagonistic interactions (signs) to the links, simulation is started in the disordered PM phase at high independence parameter $p$, with random initial orientations of spins. Then the independence parameter is decreased in small steps toward zero, and at each intermediate value of $p$, after a sufficiently long transient, the order parameters for the FM and the possible SG-like transitions are evaluated as averages over the time series of the opinion configurations. Alternatively, to check for the presence of the hysteresis loop in the first-order FM transition, simulation can be started with FM initial conditions, with all spins directed up (or down), and $p$ can be increased. The results are then averaged over $100-500$ (depending on $N$) realizations of the network and of the distribution of the signs of links.

The order parameter for the FM transition is the absolute value of the magnetization
\begin{equation}
M=\left| \left[ \langle \frac{1}{N}\sum_{j=1}^{N} \sigma_{j}\rangle_{t}\right]_{av} \right|
 \equiv \left|\left[ \langle \tilde{m} \rangle_{t} \right]_{av} \right|,
\label{M}
\end{equation}
where $\tilde{m}$ denotes a momentary value of the magnetization at a given MCSS, $\langle \cdot \rangle_{t}$ denotes the time average for a model with a given realization of the network according to $P(k)$ and with a given associated distribution of the signs of links, and $\left[ \cdot \right]_{av}$ denotes average over different realizations of the network and over different associated distributions of the signs of links. The order parameter for the SG-like transition (henceforth called the SG order parameter) is the absolute value of the overlap parameter
\cite{Binder86,Mezard87,Nishimori01},
\begin{equation}
Q=\left| \left[ \langle \frac{1}{N}\sum_{j=1}^{N} \sigma_{j}^{\alpha} \sigma_{j}^{\beta} \rangle_{t} \right]_{av}\right| 
\equiv \left| \left[ \langle \tilde{q} \rangle_{t} \right]_{av}\right|,
\label{Q}
\end{equation}
where $\alpha$, $\beta$ denote two copies (replicas) of the system simulated independently with different random initial
orientations of spins and $\tilde{q}$ is a momentary value of the overlap of their spin configurations at a given MCSS. 
In the PM phase both $M$ and $Q$ are close to zero. In the case of the FM transition both $M$ and $Q$ increase as
$p$ is decreased. In the case of the SG-like transition the SG order parameter $Q$ increases as $p$ is decreased while the magnetization $M$ remains close to zero.

The order of the FM or SG-like transition and the critical values of the independence parameter can be conveniently
determined using the respective Binder cumulants $U^{(M)}$ vs.\ $p$ and $U^{(Q)}$ vs.\ $p$ \cite{Binder97}, 
\begin{equation}
U^{(M)}=\frac{1}{2}\left[ 3-\frac{\langle \tilde{m}^{4} \rangle_{t}}{\langle \tilde{m}^{2}\rangle_{t}^{2}} \right]_{av},
\label{ULM}
\end{equation}
\begin{equation}
U^{(Q)}=\frac{1}{2}\left[ 3-\frac{\langle \tilde{q}^{4} \rangle_{t}}{\langle \tilde{q}^{2}\rangle_{t}^{2}} \right]_{av}.
\label{ULQ}
\end{equation}
In the case of the second-order FM or SG-like transition, the respective cumulants are monotonically decreasing functions of
the independence parameter: for $p\rightarrow 0$ there is $U^{(M)}\rightarrow 1$ in the FM phase and $U^{(Q)}\rightarrow 1$
in the SG phase, and for $p\rightarrow 1$ there is $U^{(M)}\rightarrow 0$, $U^{(Q)}\rightarrow 0$, respectively.
The critical values of the independence parameter  $p_{c,MC}^{(FM)}$ or $p_{c,MC}^{(SG)}$ for the FM and SG-like transitions can be determined from the intersection point of the respective Binder cumulants for models with different numbers of agents $N$ \cite{Binder97}. In the case of the first-order FM transition it is sometimes possible to observe directly the hysteresis loop, by measuring magnetization $M$ as a function of decreasing independence parameter for a model started in the PM phase to get $p_{c1,MC}^{(FM)}$, as well as as a function of increasing independence parameter for a model started in the FM phase to get $p_{c2,MC}^{(FM)}$; in order to obtain reliable results simulations of the model with the maximum number of nodes $N=10^4$ are utilized for this purpose. If the hysteresis loop is narrow the Binder cumulants again become useful. In the case of the first-order transition behavior of the cumulant $U^{(M)}$ for $p\rightarrow 0$ and $p\rightarrow 1$ is similar as in the case of the second-order transition, but close to the critical value of $p$ the cumulant exhibits negative minimum which deepens and becomes sharper with an increasing number of nodes $N$. The critical value of the independence parameter, e.g., $p_{c1,MC}^{(FM)}$ for the first-order FM transition again can be determined from the intersection point of the cumulants $U^{(M)}$ for models with different numbers of agents $N$, started in the PM phase.

\subsection{Ferromagnetic phase transition}

\label{sec:resFM}

\begin{figure}
    \includegraphics[width=0.5\linewidth]{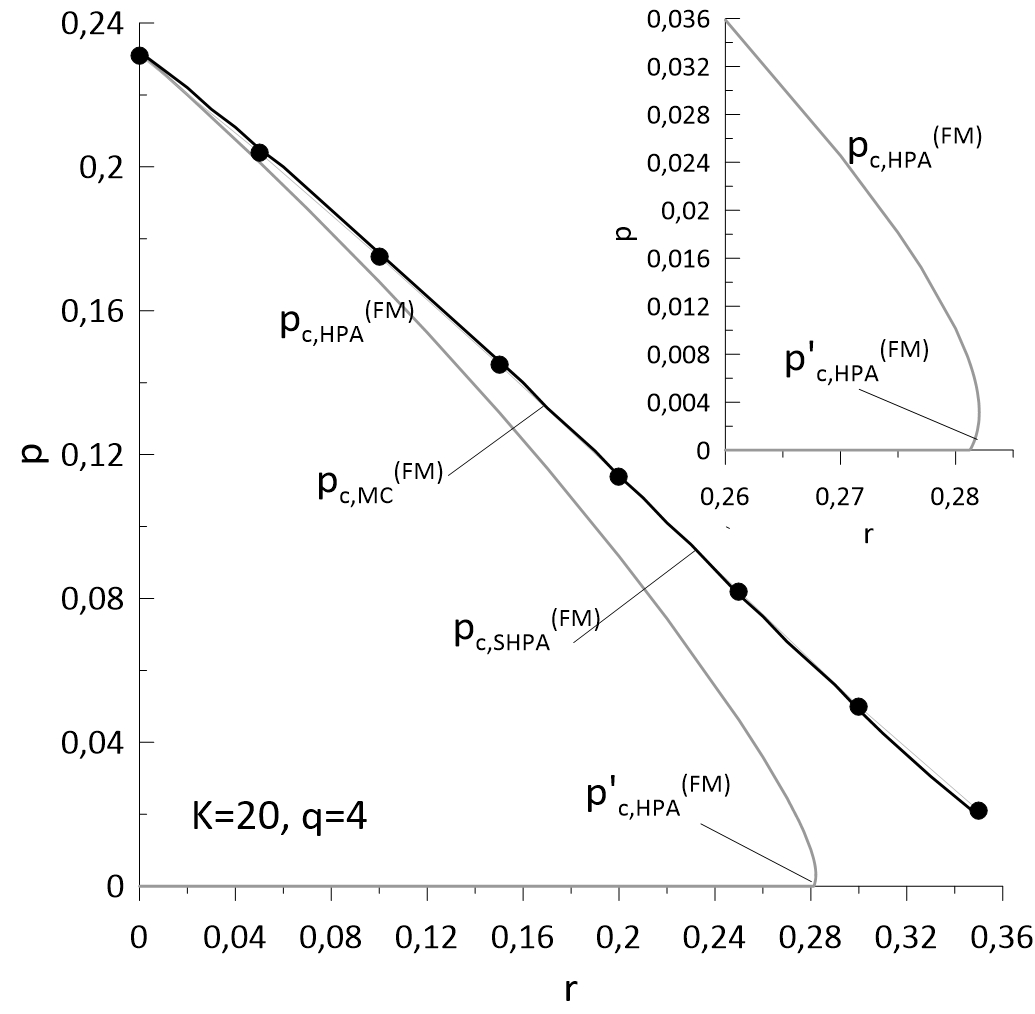}
    \caption{Phase diagram for the $q$-voter model with independence on signed RRGs with $K=20$, $q=4$. The particular critical lines for the FM transition are labeled on the diagram. Here, and in Fig.\ \ref{fig:pd_k50_q8}, \ref{fig:pd_k10_q6}, \ref{fig:pd_k10_q8} below, symbols denote critical lines obtained from MC simulations for the first-order ($\circ$) and second-order ($\bullet$) FM transition, thin gray lines are guides to the eyes. Solid lines denote critical lines predicted by the MFA (thin solid line), HPA (thick gray line) and SHPA (thick black line). Inset: meeting of the two critical lines $p_{c,HPA}^{(FM)}(r)$ and $p_{c,HPA}^{\prime (FM)}(r)$ predicted by the HPA (thick gray line) at $r=r^{\star}$.}
    \label{fig:pd_k20_q4}
\end{figure}

\begin{figure}
    \includegraphics[width=0.5\linewidth]{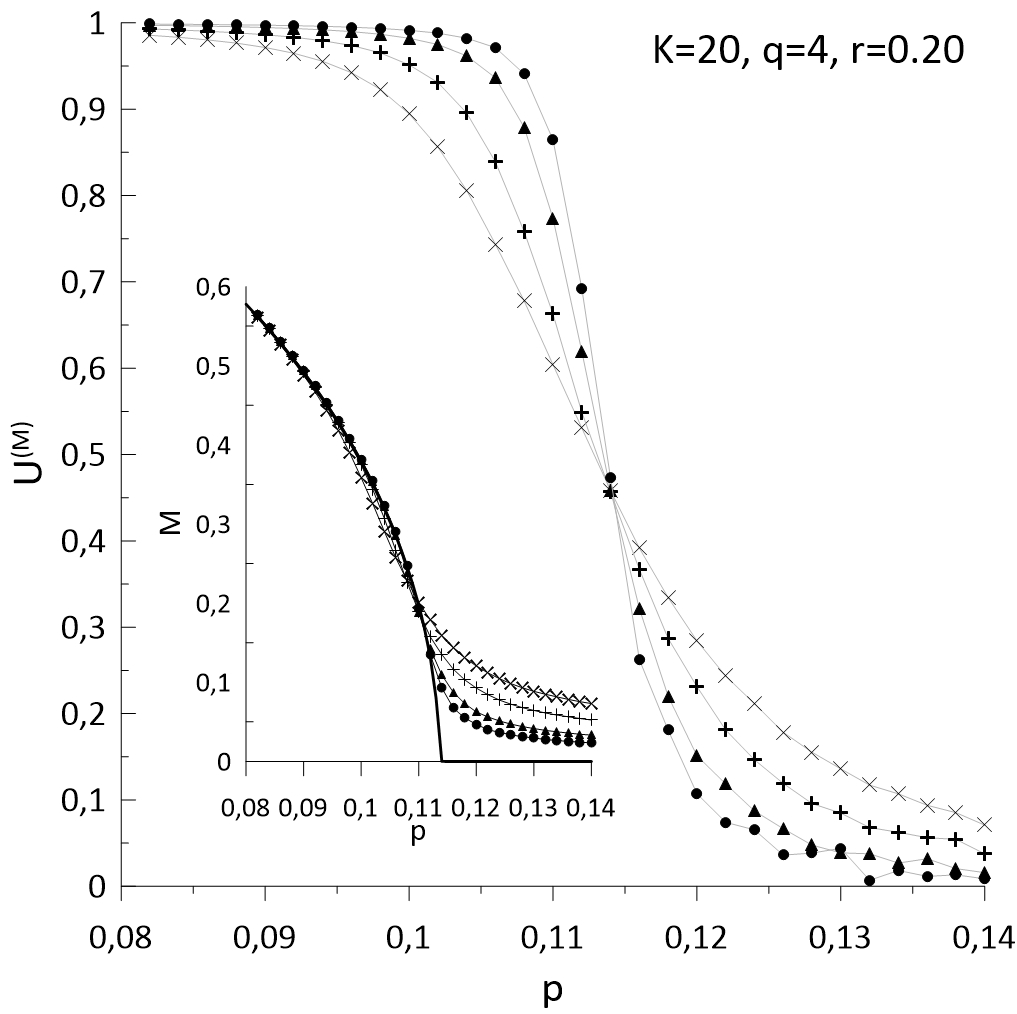}
    \caption{The Binder cumulants $U^{(M)}$ vs.\ $p$ from MC simulations of the $q$-voter model  with independence on signed RRGs with $K=20$, $q=4$, $r=0.2$ for $N=10^3$ ($\times$), $N=2\cdot 10^3$ ($+$), $N=5\cdot 10^3$ ($\blacktriangle$), $N=10^4$  ($\bullet$), gray solid lines are guides to the eyes. Inset: magnetization $M$ vs.\ $p$, symbols as above, thick black line shows predictions of the SHPA.}
    \label{fig:u4_k20_q4_p020}
\end{figure}

MC simulations of the $q$-voter model on signed RRGs and ERGs confirm that for a broad range of parameters $\langle k \rangle \ge q\ge 2$ with varying the independence parameter $p$ the FM transition occurs for a range of non-zero concentrations of the antagonistic interactions $r$. In particular, for fixed $\langle k \rangle$, $q$ the critical value of $p$ decreases to zero at $r_{max,MC}^{(FM)}< 0.5$, in agreement with Eq.\ (\ref{rmaxMF}). This situation qualitatively resembles that for the FM transition in the model for dilute SG \cite{Viana85} as well in the related nonequilibrium majority-vote \cite{Krawiecki20} and $q$-neighbor Ising models on random graphs \cite{Krawiecki21}, where the FM transition is observed for a range of non-zero fractions of the AFM interactions.

For any finite $\langle k \rangle$, $q\le 5$ ($\langle k \rangle >q$) and any $r$ the FM  transition with decreasing $p$ is second-order, as predicted by the HPA and SHPA and confirmed by MC simulations (Fig.\ \ref{fig:pd_k20_q4}, \ref{fig:u4_k20_q4_p020}). If $\langle k \rangle \gg q$, for small $r$ the critical values of the independence parameter $p_{c,HPA}^{(FM)}$, $p_{c,SHPA}^{(FM)}$ predicted from the HPA, Eq.\ (\ref{thetacpm_HPA}, \ref{pcpm_HPA}), and SHPA agree quantitatively with $p_{c,MC}^{(FM)}$ obtained from MC simulations (Fig.\ \ref{fig:pd_k20_q4}). For moderate $r$ the critical values $p_{c,HPA}^{(FM)}$ predicted by the HPA become significantly underestimated. Besides, for $r$ slightly below $r^{\star}$, Eq.\ (\ref{rstar_HPA}), the HPA incorrectly predicts the transition from the PM to the FM phase with decreasing $p$ at $p_{c,HPA}^{\prime (FM)}$ which is not observed in simulations (Fig.\ \ref{fig:pd_k20_q4}). Both critical curves $p_{c,HPA}^{(FM)} (r)$, $p_{c,HPA}^{\prime (FM)} (r)$ merge at $r=r^{\star}$ marking a border of the range of $r$ where the FM transition can occur, noticeably below the bordering value $r_{max,MC}^{(FM)}$ estimated from MC simulations (Fig.\ \ref{fig:pd_k20_q4}). In contrast, the critical values $p_{c,SHPA}^{(FM)}$ predicted by the SHPA agree quantitatively with $p_{c,MC}^{(FM)}$ obtained from MC simulations for a whole range of $r$ where the FM transition appears (Fig.\ \ref{fig:pd_k20_q4}), and for fixed $r$ the dependence of the magnetization on $p$ below the transition point is also correctly reproduced (Fig.\ \ref{fig:u4_k20_q4_p020}).

\begin{figure}[h]
    \includegraphics[width=0.5\linewidth]{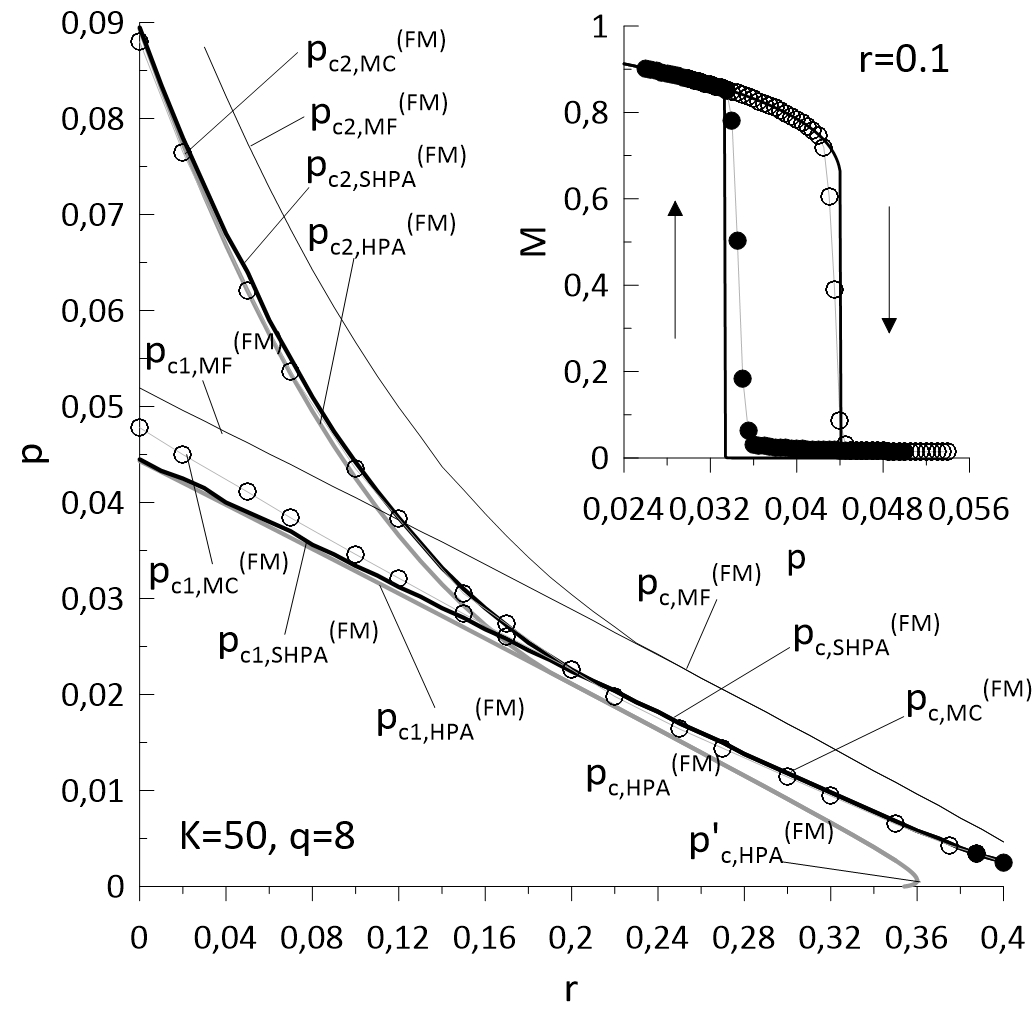}
    \caption{As in Fig.\ \ref{fig:pd_k20_q4}, but for $K=50$, $q=8$. Inset: magnetization $M$ vs.\ $p$ from MC simulations of the model with $r=0.1$ started with PM conditions and decreasing $p$ ($\bullet$) as well as with FM conditions and increasing $p$ ($\circ$), thick black line shows predictions of the SHPA}.
    \label{fig:pd_k50_q8}
\end{figure}

\begin{figure}[h]
    \includegraphics[width=0.5\linewidth]{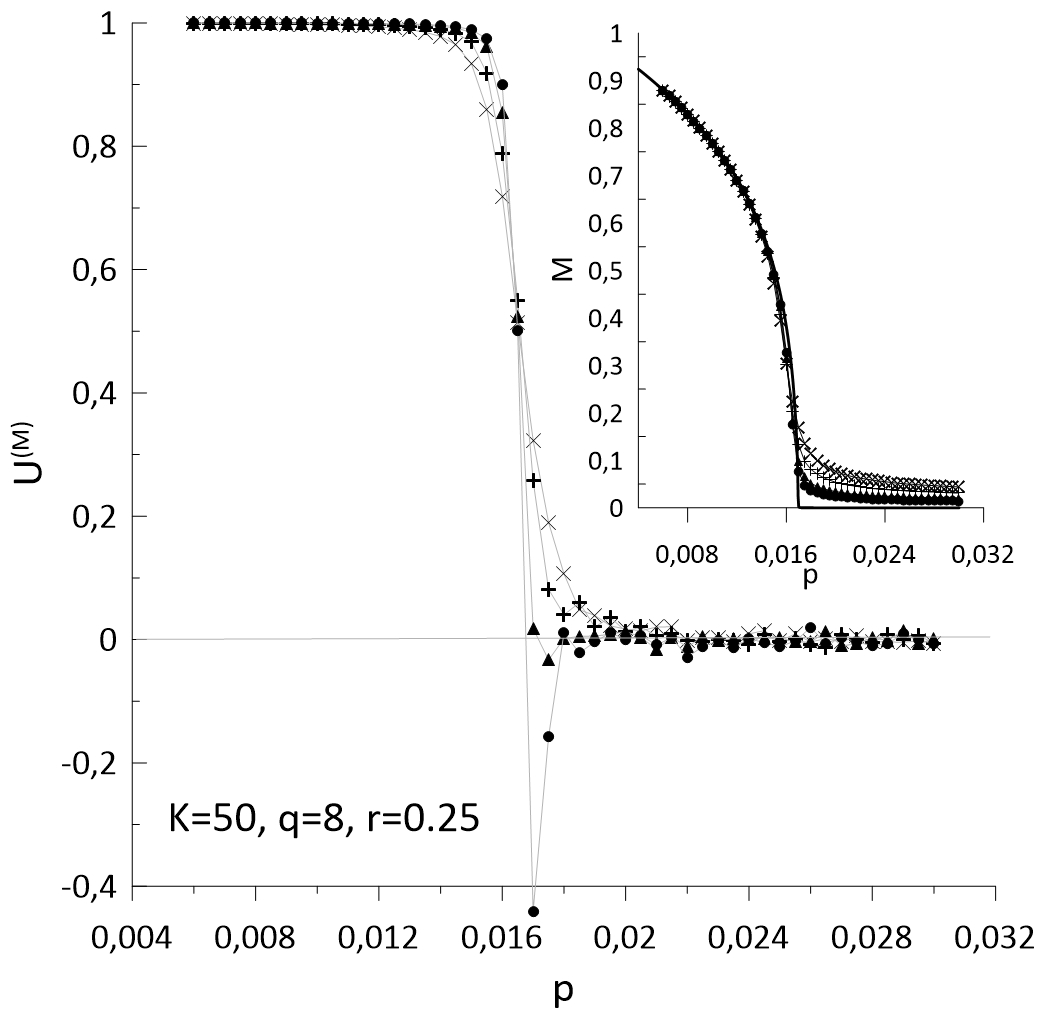}
    \caption{As in Fig.\ \ref{fig:u4_k20_q4_p020} but for $K=50$, $q=8$, $r=0.25$.}
    \label{fig:u4_k50_q8_p025}
\end{figure}

For $\langle k \rangle \ge 6$ the phase diagram on the $p$ vs $r$ plane for the FM transition is more complex. If $\langle k \rangle \gg q$ the properties of the FM transition with fixed $r$ and varying $p$ are qualitatively reproduced by the MFA (Fig.\ \ref{fig:pd_k50_q8}): for small $r$ it is discontinuous with the hysteresis loop, for larger $r$ it is continuous, and the critical lines for the first- and second-order transitions meet in the TCP. For small $r$ the critical values of the independence parameter $p_{c1,HPA}^{(FM)}$, $p_{c2,HPA}^{(FM)}$ and $p_{c1,SHPA}^{(FM)}$, $p_{c2,SHPA}^{(FM)}$ for the first-order transition as well as $p_{c,HPA}^{(FM)}$, $p_{c,SHPA}^{(FM)}$ for the second-order transition, predicted by the HPA and SHPA, respectively, agree quantitatively with those obtained from MC simulations, $p_{c1,MC}^{(FM)}$, $p_{c2,MC}^{(FM)}$ as well as $p_{c,MC}^{(FM)}$ for the first- and second-order transition, respectively; thus, the width of the hysteresis loop in the case of the discontinuous transition is also predicted correctly by both kinds of the PA (Fig.\ \ref{fig:pd_k50_q8}). The location of the TCP $\left( r_{TCP,MC}^{(FM)}, p_{TCP,MC}^{(FM)}\right)$ obtained from MC simulations, in which the width of the hysteresis loop decreases to zero, is also predicted correctly by the HPA and SHPA (Fig.\ \ref{fig:pd_k50_q8}). However, for a range of $r$ above $r_{TCP,MC}^{(FM)}$
the FM transition observed in MC simulations is still first-order (Fig.\ \ref{fig:pd_k50_q8}): although the hysteresis loop is not observed directly, the Binder cumulants $U^{(M)}$ for different $N$ cross at one point corresponding to the critical value $p_{c,MC}^{(FM)}$ and exhibit negative minima as functions of $p$ which become deeper with an increasing number of nodes (Fig.\ \ref{fig:u4_k50_q8_p025}). It is possible that the presence of this minimum is a finite-size effect that will disappear again in the thermodynamic limit, the more that the increase of the magnetization with decreasing $p$ is correctly predicted by the SHPA and typical for the second-order transition (Fig.\ \ref{fig:u4_k50_q8_p025}, inset), but simulations of the model with $N$ large enough are beyond our capabilities. Only for still higher values of $r$ does the FM transition observed in MC simulations become continuous, as predicted by both kinds of the PA (Fig.\ \ref{fig:u4_k50_q8_p025}). For moderate $r$, in particular, in the case of the second-order FM transition, predictions of the HPA again deviate from the results of MC simulations in a similar way as in the above-mentioned case with $\langle k \rangle=20$, $q=4$. In contrast, the critical values $p_{c,SHPA}^{(FM)}$ predicted by the SHPA agree quantitatively with $p_{c,MC}^{(FM)}$ obtained from MC simulations (Fig.\ \ref{fig:u4_k50_q8_p025}).

\begin{figure}[h]
    \includegraphics[width=0.5\linewidth]{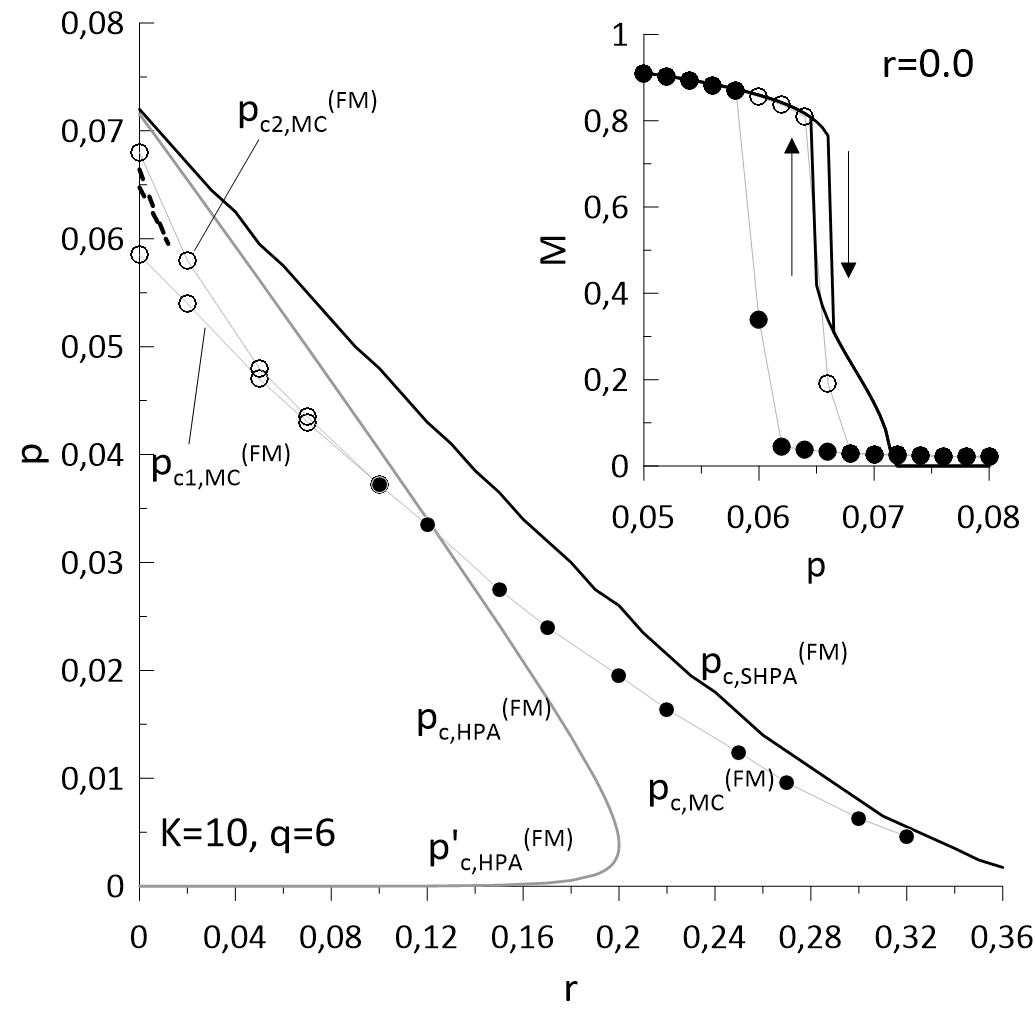}
    \caption{As in Fig.\ \ref{fig:pd_k20_q4}, but for $K=10$, $q=6$; dashed lines denote the lower and upper critical lines for the additional discontinuous FM transition predicted by the SHPA, occurring for $p<p_{c,SHPA}^{(FM)}$, i.e., below the continuous FM transition, as shown in the inset. Inset: as in Fig.\ \ref{fig:pd_k50_q8}, for the model with $r=0.0$.}
    \label{fig:pd_k10_q6}
\end{figure}

\begin{figure}[h]
    \includegraphics[width=0.5\linewidth]{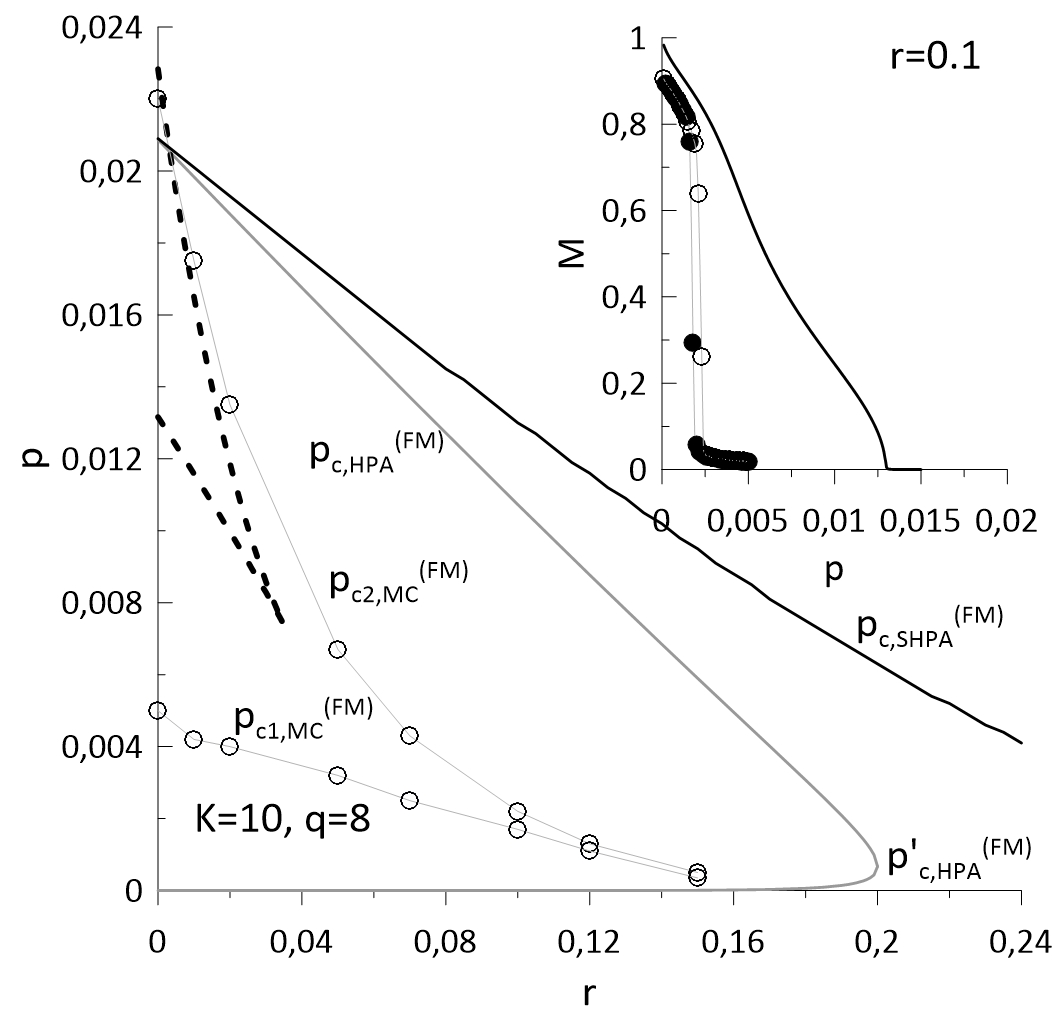}
    \caption{As in Fig.\ \ref{fig:pd_k20_q4}, but for $K=10$, $q=8$; dashed lines as in Fig.\ \ref{fig:pd_k10_q6}. Inset: as in Fig.\ \ref{fig:pd_k50_q8}, for the model with $r=0.1$.}
    \label{fig:pd_k10_q8}
\end{figure}

If $\langle k \rangle$ and $q$ are comparable, the agreement between the results of MC simulations and all theoretical predictions under consideration is much worse (Fig.\ \ref{fig:pd_k10_q6}, \ref{fig:pd_k10_q8}). Such disagreement appears even in the case of the model with $r=0$ \cite{Jedrzejewski17,Gradowski20,Jedrzejewski22}. It should be mentioned that in order to study this case, $\langle k \rangle$ must be diminished rather than $q$ increased, since the rise of $q$ leads to the decrease of the critical value of $p$ for the occurrence of the FM transition, cf.\ Eq.\ (\ref{pcMFAFM0}), and for such small level of internal noise very long MC simulations are required to obtain reliable dependence of the magnetization on the independence parameter. As a result, for small $\langle k \rangle$ predictions of the MFA are useless. Concerning the HPA and SHPA, for small $r$ their predictions are even qualitatively incorrect: for fixed $r$ and decreasing $p$ both theories predict the occurrence of the second-order FM transition, while MC simulations reveal the first-order FM transition with the hysteresis loop (Fig.\ \ref{fig:pd_k10_q6}, \ref{fig:pd_k10_q8}). Moreover, according to the SHPA (as well as the HPA) at $p< p_{c,SHPA}^{(FM)}$ another discontinuous phase transition can appear between two FM phases with low and high magnetization (Fig.\ \ref{fig:pd_k10_q6}). This results from a sequence of bifurcations of the system of equations (\ref{dcdt_SHPA},\ref{depuudt_SHPA} - \ref{demdddt_SHPA}) with decreasing $p$, first the supercritical pitchfork bifurcation at $p= p_{c,SHPA}^{(FM)}$, leading to the second-order FM transition to the FM phase with low magnetization, then two saddle-node bifurcations leading to bistability between two FM phases with low and high magnetization and eventually to the loss of stability of the former FM phase. The range of parameters $r$, $p$ for the occurrence of this additional FM transition is much narrower than that for the first-order FM transition observed in MC simulations. For moderate $r$ MC simulations show that for smaller $q$ the FM transition can become second-order (Fig.\ \ref{fig:pd_k10_q6}), as predicted by the HPA and SHPA, while for larger $q$ it remains first-order (Fig.\ \ref{fig:pd_k10_q8}). In the former case, predictions of the HPA again deviate from the results of MC simulations in a similar way as in the above-mentioned cases with $\langle k \rangle \gg q$, while the critical values $p_{c,SHPA}^{(FM)}$ predicted by the SHPA approach $p_{c,MC}^{(FM)}$ obtained from simulations. In the latter case the order of the FM transition predicted by both HPA and SHPA is incorrect and the critical values of the independence parameter are significantly overestimated.  

\subsection{Spin-glass-like transition}

\label{sec:resSG}

\begin{figure}[h]
    \includegraphics[width=0.5\linewidth]{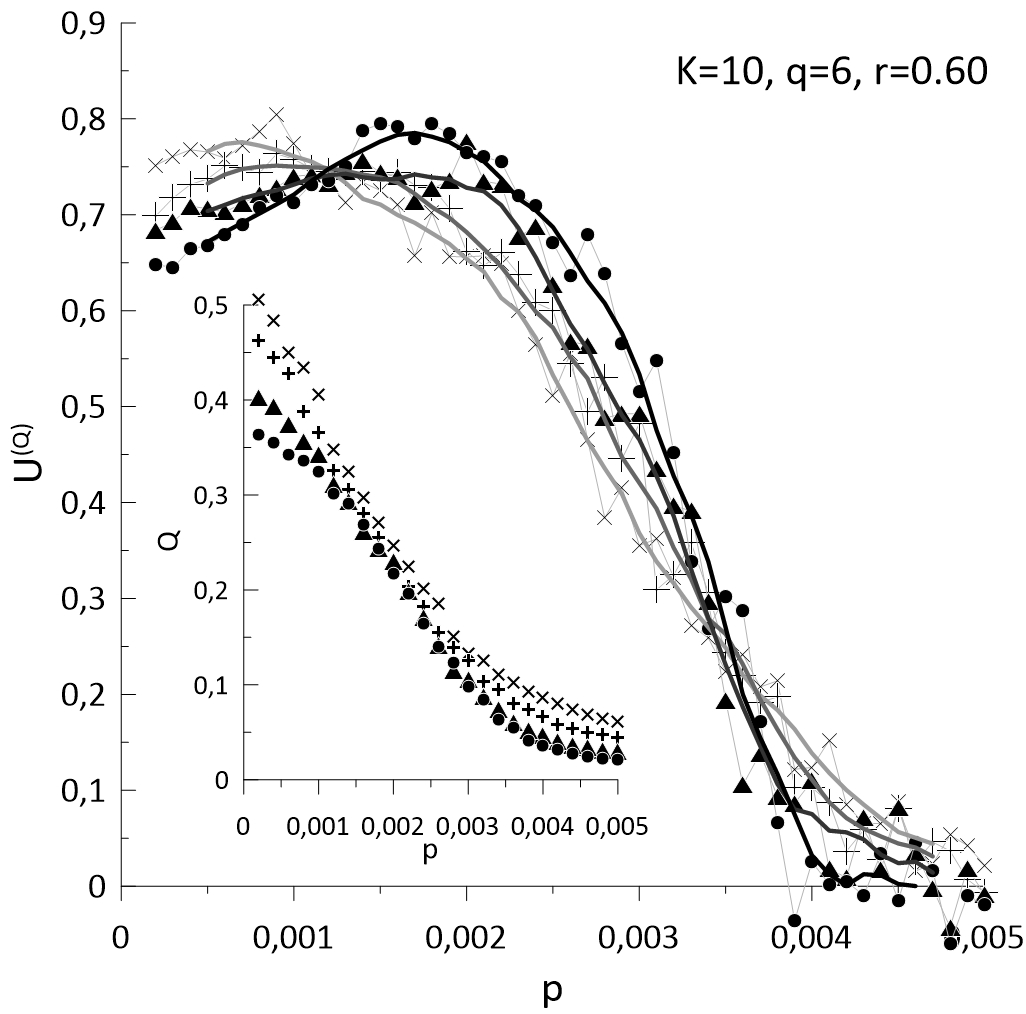}
    \caption{The Binder cumulants $U^{(Q)}$ vs.\ $p$ from MC simulations of the $q$-voter model  with independence on signed RRGs with $K=10$, $q=6$, $r=0.6$ for $N=10^3$ ($\times$), $N=2\cdot 10^3$ ($+$), $N=5\cdot 10^3$ ($\blacktriangle$), $N=10^4$  ($\bullet$), thin gray solid lines are guides to the eyes, thick solid lines are running averages with the window width equal to 7 points (from light gray for $N=10^3$ to black for $N=10^4$). Inset: the SG order parameter $Q$ vs.\ $p$, symbols as above.}
    \label{fig:u4q_k10_q6_p060}
\end{figure}

\begin{figure}[h]
    \includegraphics[width=0.5\linewidth]{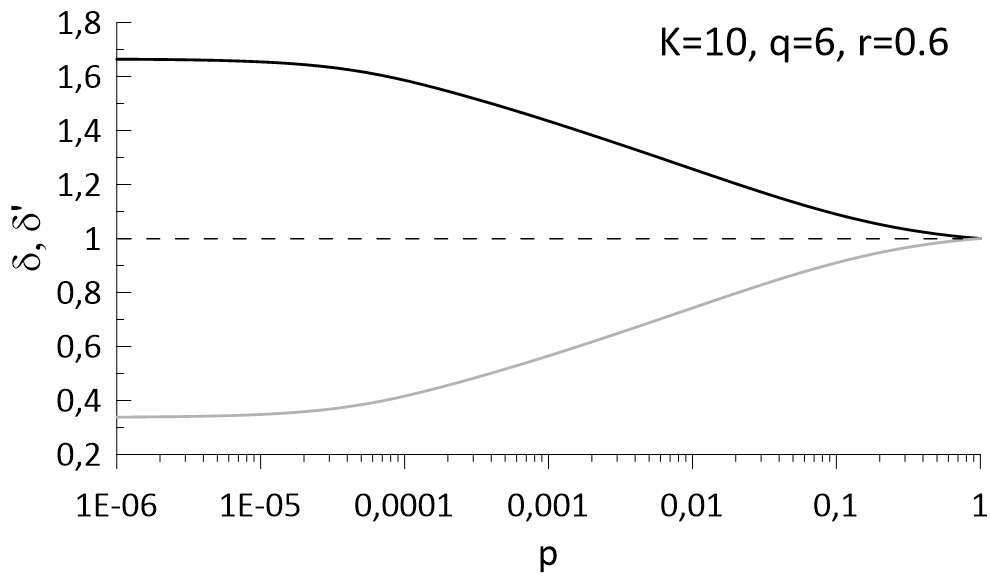}
    \caption{Relative concentrations $\delta=e_{\uparrow,\uparrow}^{(+)}/[c^2 (1-r)]$, $\delta'= e_{\uparrow,\uparrow}^{(-)}/(c^2 r)$ of active and inactive links vs.\ $p$ predicted by the SHPA for the $q$-voter model with independence on signed RRGs with $K=10$, $q=6$, $r=0.6$, where MC simulations reveal the SG-like transition for decreasing independence parameter.}
    \label{fig:conc_rel}
\end{figure}

In the nonequilibrium majority-vote \cite{Krawiecki20} and $q$-neighbor Ising model \cite{Krawiecki21} on signed networks with fixed $r$ and decreasing level of internal noise numerical evidence was found for the occurrence of the SG-like transition, recognized by the increase of the SG order parameter $Q$ with the magnetization $M$ remaining zero. The resulting phase diagram for the above-mentioned models qualitatively resembles that of the equilibrium dilute SG model \cite{Viana85}, with the FM and SG-like transitions occurring for small and large fractions of the antagonistic interactions, respectively, separated by a TCP, and with the critical value  of the parameter measuring the intensity of the internal noise (e.g., the temperature in the case of the $q$-neighbor Ising model) for the SG-like transition independent of $r$. The SG-like transition cannot be predicted theoretically using methods of Sec.\ \ref{sec:theory}, only a little hint of it can be obtained from the HPA and SHPA. 

It turns out that in the $q$-voter model on signed networks considered in this paper, it is particularly difficult to observe the SG-like transition in MC simulations. Exemplary numerical results indicating its occurrence in the model on an RRG with $K=10$, $q=6$, $r=0.6$, where the FM transition does not appear, are shown in Fig.\ \ref{fig:u4q_k10_q6_p060}. The SG order parameter $Q$ increases monotonically for $p\rightarrow 0$, although its values do not saturate and slightly decrease with increasing $N$ which can raise doubts about the occurrence of the SG-like transition in the thermodynamic limit. Since the transition is observed for very small $p$ the obtained curves $U^{Q}$ vs.\ $p$ exhibit strong fluctuations despite averaging the results as described in Sec.\ \ref{sec:resMET}; only after smoothing them the Binder cumulants for different $N$ seem to decrease monotonically with $p$ and cross at one point corresponding to the critical value $p_{c,MC}^{(SG)}= 0.0037\pm 0.0002$. These results suggest the appearance of the second-order SG-like transition in the model under study with large $r$, with the critical line meeting in a TCP with that for the second-order FM transition seen in Fig.\ \ref{fig:pd_k10_q6}, as in dilute SG models \cite{Viana85}. 

It is interesting to note that the obtained value $p_{c,MC}^{(SG)}$ is close to the value $p_{c,HPA}^{(FM)}(r^{\star})= p_{c,HPA}^{\prime (FM)}(r^{\star}) = 0.00382618\ldots$ of the independence parameter at the cusp of the region of stability of the FM phase predicted by the HPA (Fig.\ \ref{fig:pd_k10_q6}). A similar coincidence was observed also in the case of the $q$-neighbor Ising model on signed networks \cite{Krawiecki21}, although its origin is unclear, the more that the presence of the additional critical line $p_{c,HPA}^{\prime (FM)}(r)$ as well as the cusp at $r=r^{\star}$ are only due to approximations made in the derivation of the HPA (Sec.\ \ref{sec:theory_hpa}). It is also noteworthy to mention that for large $r$, where the MC simulations suggest the appearance of the SG-like transition, the HPA predicts that the PM point remains stable for $p\rightarrow 0$ but 
the corresponding concentration of active links $b=\theta$ decreases significantly which suggests that some correlation between orientations of interacting spins appears. A similar prediction is also made by the SHPA. It predicts that for large $r$ and $p\rightarrow 0$ the links corresponding to reinforcing interactions connect mostly nodes occupied by spins with the same orientations, while these corresponding to antagonistic interactions connect mostly nodes occupied by spins with opposite orientations, which suggests the appearance of short-range ordering of spins. For example, in the above-mentioned case of the model on RRG with $K=10$, $q=6$, $r=0.6$, for $p\rightarrow 0$ the ratio $\delta=e_{\uparrow,\uparrow}^{(+)}/[c^2 (1-r)]$ (where the denominator corresponds to the concentration of positive links connecting spins with orientation up if the positive and negative links are distributed randomly between pairs of nodes, as in the PM phase; note that also in the SG phase $c=1/2$) significantly increases, while $\delta'= e_{\uparrow,\uparrow}^{(-)}/(c^2 r)$ similarly decreases (Fig.\ \ref{fig:conc_rel}). Substantial changes of $\delta$, $\delta'$ are predicted only for very small $p$ which suggests that the critical value of the independence parameter for the SG-like transition is small which hampers its observation in MC simulations. 

\section{Summary and conclusions}. 

\label{sec:conclusions}

In this paper, the $q$-voter model with independence on signed random graphs was studied by MC simulations and theoretically in the MF approximation as well as using two versions of the PA, the HPA and SHPA. The independence, measured by the parameter $p$, corresponds to internal noise in the model, and the signed networks exhibit quenched disorder due to the appearance of negative links corresponding to antagonistic interactions with probability $r$. The latter interactions prefer opposite opinions (orientations) of the interacting agents (spins), which is reflected in the modified rule for the update of agents' opinions; thus, the model under study qualitatively resembles the equilibrium  dilute SG model with a fraction $r$ of the AFM interactions \cite{Viana85} and previously studied nonequilibrium majority-vote and $q$-neighbor Ising model \cite{Krawiecki20,Krawiecki21} on signed networks. Indeed, also the FM transition with fixed $r$ and varying $p$ observed in the $q$-voter model with independence on signed networks with a finite mean degree of nodes $\langle k \rangle$ qualitatively resembles that in the above-mentioned nonequilibrium models: for the size of the $q$-neighborhood $q\le 5$ it is second-order for any $r$; for $q\ge 6$ it is first-order with the hysteresis loop for small $r$ and can become second-order for larger $r$; and it disappears above a certain level of quenched disorder $r<1$. Besides, numerical evidence was found for the occurrence of the SG-like transition in the model under study with a large fraction $r$ of antagonistic interactions, again in analogy with the above-mentioned models in Ref.\ \cite{Viana85,Krawiecki20,Krawiecki21}. 

Theoretical predictions of the MFA and HPA concerning the FM transition in the model under study exhibit similar discrepancies with results of MC simulations as in the case of the majority-vote and $q$-neighbor Ising models \cite{Krawiecki20,Krawiecki21}. The MFA yields quantitatively correct predictions concerning the order of the FM transition and the critical value(s) of the independence parameter $p$ for the model on complete graphs with $\langle k \rangle \rightarrow \infty$, as expected. Predictions of the HPA, which to a large extent can be obtained analytically, are quantitatively correct for the model on networks with a finite mean degree of nodes $\langle k\rangle$ substantially larger than $q$, and for small fractions of antagonistic interactions $r$. As $r$ is increased, the predicted critical values $p_{c,HPA}^{(FM)}$ of the independence parameter deviate from these obtained from MC simulations, and for a certain range of $r$ the theory predicts also additional transition from the FM back to the PM phase with $p\rightarrow 0$ which is not observed in simulations. Eventually, for the model on networks with $\langle k \rangle$ comparable with $q$, predictions of the HPA become even qualitatively incorrect. 

The main theoretical result of this paper is the formulation of the SHPA and its application to the $q$-voter model under study in Sec.\ \ref{sec:theory_shpa}. This kind of PA, valid for models on signed networks, distinguishes between different kinds (classes) of links with a given sign connecting spins with given orientations so that the macroscopic quantities are concentrations of spins with orientation up and the above-mentioned links. Derivation of the equations of motion for the macroscopic quantities in the SHPA resembles that in the PA for the $q$-voter models with quenched disorder on networks \cite{Jedrzejewski22}; however, in the latter case, the quenched disorder is connected with the presence of two kinds of agents in the model, differing by the rules for the opinion update, rather than two kinds of reinforcing and antagonistic interactions, which eventually leads to a different formulation of the theory. For the model under study on networks with a finite mean degree of nodes $\langle k\rangle$ substantially larger than $q$ the critical values $p_{c,SHPA}^{(FM)}$ of the independence parameter predicted by the SHPA show good quantitative agreement with these obtained from MC simulations for a whole range of $r$ where the FM transition occurs, and the unobserved transition to the PA phase with $p\rightarrow 0$ is not predicted. The SHPA offers also heuristic insight into the origin of the SG-like phase observed in the model under study for large $r$, since it suggests that in this case, the concentration of active links decreases for $p\rightarrow 0$ and the correlation between the signs of links and the orientations of spins in the connected nodes increases in comparison with the PM phase. Unfortunately, for the model on networks with $\langle k \rangle$ comparable with $q$, predictions of the SHPA concerning the FM transition again become even qualitatively incorrect. 

Similar improvement of theoretical predictions concerning the FM transition in comparison with the HPA can be expected due to applying the SHPA to the related majority-vote and $q$-neighbor Ising models on signed networks. It is also worth noting that one of the assumptions of all theoretical approaches in Sec.\ \ref{sec:theory} is that all nodes are statistically equivalent, thus the macroscopic quantity of interest is the total concentration $c$ of spins with orientation up. However, even in the case of a signed network in the form of a RRG, the nodes differ by the number $l$ of attached antagonistic directed links which obeys binomial distribution $B_{K,l}(r)$. Taking this heterogeneity into account may lead to more advanced theoretical approaches for the $q$-voter and related models on signed networks. The above-mentioned issues are left for future investigation.


\begin{thebibliography}{0}%
\makeatletter
\providecommand \@ifxundefined [1]{%
 \@ifx{#1\undefined}
}%
\providecommand \@ifnum [1]{%
 \ifnum #1\expandafter \@firstoftwo
 \else \expandafter \@secondoftwo
 \fi
}%
\providecommand \@ifx [1]{%
 \ifx #1\expandafter \@firstoftwo
 \else \expandafter \@secondoftwo
 \fi
}%
\providecommand \natexlab [1]{#1}%
\providecommand \enquote  [1]{``#1''}%
\providecommand \bibnamefont  [1]{#1}%
\providecommand \bibfnamefont [1]{#1}%
\providecommand \citenamefont [1]{#1}%
\providecommand \href@noop [0]{\@secondoftwo}%
\providecommand \href [0]{\begingroup \@sanitize@url \@href}%
\providecommand \@href[1]{\@@startlink{#1}\@@href}%
\providecommand \@@href[1]{\endgroup#1\@@endlink}%
\providecommand \@sanitize@url [0]{\catcode `\\12\catcode `\$12\catcode
  `\&12\catcode `\#12\catcode `\^12\catcode `\_12\catcode `\%12\relax}%
\providecommand \@@startlink[1]{}%
\providecommand \@@endlink[0]{}%
\providecommand \url  [0]{\begingroup\@sanitize@url \@url }%
\providecommand \@url [1]{\endgroup\@href {#1}{\urlprefix }}%
\providecommand \urlprefix  [0]{URL }%
\providecommand \Eprint [0]{\href }%
\providecommand \doibase [0]{https://doi.org/}%
\providecommand \selectlanguage [0]{\@gobble}%
\providecommand \bibinfo  [0]{\@secondoftwo}%
\providecommand \bibfield  [0]{\@secondoftwo}%
\providecommand \translation [1]{[#1]}%
\providecommand \BibitemOpen [0]{}%
\providecommand \bibitemStop [0]{}%
\providecommand \bibitemNoStop [0]{.\EOS\space}%
\providecommand \EOS [0]{\spacefactor3000\relax}%
\providecommand \BibitemShut  [1]{\csname bibitem#1\endcsname}%
\let\auto@bib@innerbib\@empty
\end{thebibliography}%


\begin{thebibliography}{00}

\bibitem{Castellano09}
C.\ Castellano, S.\ Fortunato, V.\ Loreto, 
Statistical physics of social dynamics.
Rev.\ Mod.\ Phys.\ 81, 591 (2009).


\bibitem{Clifford73}
P.\ Clifford, and A. Sudbury, 
A model for spatial conflict,
Biometrika 60, 581 (1973). 

\bibitem{Holley75}
R.\ Holley and T. Liggett, 
Ergodic theorems for weakly interacting infinite systems and the voter model,
Ann.\ Probab.\ 3, 643 (1975).

\bibitem{Sood05}
V.\ Sood, S.\ Redner, 
Voter model on heterogeneous graphs, 
Phys.\ Rev.\ Lett.\ 94, 178701 (2005).

\bibitem{Sood08}
V.\ Sood, T.\ Antal, S.\ Redner, 
Voter models on heterogeneous networks, 
Phys.\ Rev.\ E 77, 041121 (2008).

\bibitem{Vazquez08}
F.\ Vazquez, V.\ M.\ Egu\'{ı}luz, 
Analytical solution of the voter model on uncorrelated networks, 
New J.\ Phys.\ 10, 063011 (2008).

\bibitem{Pugliese09}
E.\ Pugliese, C.\ Castellano, 
Heterogeneous pair approximation for voter models on networks, 
Europhys.\ Lett.\ 88, 58004 (2009).


\bibitem{Granovsky95}
B.\ L.\ Granovsky, N.\ Madras,
The noisy voter model,
Stochastic Processes and their Applications 55, 23 (1995).

\bibitem{Carro16}
A.\ Carro, R.\ Toral and M.\ San Miguel,
The noisy voter model on complex networks,
Sci Rep 6, 24775 (2016).

\bibitem{Peralta18a}
A.\ F.\ Peralta, A.\ Carro, M.\ San Miguel, and R.\ Toral,
Stochastic pair approximation treatment of the noisy voter model,
New J.\ Phys.\ 20, 103045 (2018).


\bibitem{Castellano09a}
C.\ Castellano, M.\ A.\ Mu\~noz, and R.\ Pastor-Satorras, 
Nonlinear $q$-voter model,
Phys.\ Rev.\ E 80 (2009) 041129.

\bibitem{Nyczka12}
P.\ Nyczka, K.\ Sznajd-Weron and J.\ Cis\char32 lo, 
Phase transitions in the $q$-voter model with two types of stochastic driving,
Phys.\ Rev.\ E 86, 011105 (2012).

\bibitem{Moretti13}
P.\ Moretti, S.\ Liu, C.\ Castellano, R.\ Pastor-Satorras,
Mean-field analysis of the $q$-voter model on networks,
J.\ Stat.\ Phys.\  151, 113 (2013).

\bibitem{Chmiel15}
A.\ Chmiel and K.\ Sznajd-Weron, 
Phase transitions in the $q$-voter model with noise on a duplex clique,
Phys.\ Rev.\ E 92, 052812 (2015).

\bibitem{Jedrzejewski17}
A.\ J\c{e}drzejewski, 
Pair approximation for the $q$-voter model with independence on complex networks,
Phys.\ Rev.\ E 95, 012307 (2017).

\bibitem{Peralta18}
A.\ F.\ Peralta, A.\ Carro, M.\ San Miguel, and R.\ Toral,
Analytical and numerical study of the non-linear noisy voter model on complex networks,
Chaos 28, 075516 (2018).

\bibitem{Jedrzejewski22}
A.\ J\c{e}drzejewski and K.\ Sznajd-Weron,
Pair approximation for the 
$q$-voter models with quenched disorder on networks,
Phys.\ Rev.\ E 105, 064306 (2022).

\bibitem{Vieira18}
A.\ Vieira and C.\ Anteneodo,
Threshold $q$-voter model,
Phys.\ Rev.\ E 97, 052106 (2018).

\bibitem{Vieira20}
A.\ R. Vieira, A.\ F.\ Peralta, R.\ Toral, M.\ San Miguel, and C.\ Anteneodo,
Pair approximation for the noisy threshold $q$-voter model,
Phys.\ Rev.\ E 101, 052131 (2020).

\bibitem{Gradowski20}
T.\ Gradowski, A.\ Krawiecki,
Pair approximation for the $q$-voter model with independence on multiplex networks,
Phys.\ Rev. E 102, 022314 (2020).

\bibitem{Nowak21}
B.\ Nowak, B.\ Stoń, K.\ Sznajd-Weron,
Discontinuous phase transitions in the multi-state noisy $q$-voter model: quenched vs.\ annealed disorder,
Sci.\ Rep.\ 11, 6098 (2021).


\bibitem{Oliveira92}
M.\ J.\ Oliveira,
Isotropic majority-vote model on a square lattice,
J.\ Stat.\ Phys.\ 66, 273 (1992).

\bibitem{Chen15}
Hanshuang Chen, Chuansheng Shen, Gang He, Haifeng Zhang, and Zhonghuai Hou,
Critical noise of majority-vote model on complex networks,
Phys.\ Rev.\ E 91, 022816 (2015).

\bibitem{Chen17}
H.\ Chen, C.\ Shen, H.\ Zhang, G.\  Li, Z.\ Hou, and J.\ Kurths, 
First-order phase transition in a majority-vote model with inertia,
Phys.\ Rev.\ E 95, 042304 (2017).

\bibitem{Nowak20}
B.\ Nowak and K.\ Sznajd-Weron,
Symmetrical threshold model with independence on random graphs,
Phys.\ Rev.\ E 101, 052316 (2020).

\bibitem{Chen20}
Hanshuang Chen, Shuang Wang, Chuansheng Shen, Haifeng Zhang, and G.\ Bianconi, 
Non-markovian majority-vote model,
Phys.\ Rev.\ E 102, 062311 (2020).

\bibitem{Kim21}
Minsuk Kim and Soon-Hyung Yook,
Majority-vote model with degree-weighted influence on complex networks,
Phys.\ Rev.\ E 103, 022302 (2021).


\bibitem{Jedrzejewski15}
A. J\c{e}drzejewski, A.\ Chmiel, K.\ Sznajd-Weron, 
Oscillating hysteresis in the $q$-neighbor Ising model,
Phys.\ Rev.\ E 92, 052105 (2015).

\bibitem{Park17}
J.-M.\ Park and J.\ D.\ Noh, 
Tricritical behavior of nonequilibrium Ising spins in fluctuating environments,
Phys.\ Rev.\ E 95, 042106 (2017).

\bibitem{Chmiel17}
A.\ Chmiel, J.\ Sienkiewicz, and K.\ Sznajd-Weron,
Tricriticality in the $q$-neighbor Ising model on a partially duplex clique,
Phys.\ Rev.\ E 96, 062137 (2017). 

\bibitem{Chmiel18}
A.\ Chmiel, T.\ Gradowski and A.\ Krawiecki,
$q$-neighbor Ising model on random networks,
Int.\ J.\ Modern Phys.\ C 29, 1850041 (2018).


\bibitem{Albert02}
R.\ Albert and A.-L.\ Barab\'asi,
Statistical mechanics of complex networks,
Rev.\ Mod.\ Phys.\  74, 47 (2002).

\bibitem{Dorogovtsev08}
S.\ N.\ Dorogovtsev,  A.\ V.\ Goltsev, and J.\ F.\ F.\ Mendes,
Critical phenomena in complex networks,
Rev.\ Mod.\ Phys.\ 80, 1276 (2008).


\bibitem{Krawiecki20}
A.\ Krawiecki, 
Ferromagnetic and spin-glass-like transition
in the majority-vote model on complete and random
graphs,
Eur.\ Phys.\ J.\ B 93, 176 (2020)

\bibitem{Krawiecki21}
A.\ Krawiecki,
Ferromagnetic and spin-glass like transition in the
$q$-neighbor Ising model on random graphs,
Eur.\ Phys.\ J.\ B  94, 73 (2021).

\bibitem{Baron21}
J.\ W.\ Baron,
Persistent individual bias in a voter model with quenched disorder,
Phys.\ Rev.\ E 103, 052309 (2021).

\bibitem{Baron21a}
J.\ W.\ Baron,
Consensus, polarization, and coexistence in a continuous opinion dynamics model with quenched disorder,
Phys.\ Rev.\ E 104, 044309 (2021).


\bibitem{Sherrington75}
D.\ Sherrington, S.\ Kirkpatrick,
Solvable model of a spin-glass,
Phys.\ Rev.\ Lett.\ 35, 1792 (1975).

\bibitem{Binder86}
K.\ Binder, A.\ P.\ Young, 
Spin glasses: Experimental facts, theoretical concepts, and open questions,
Rev.\ Mod.\ Phys.\ 58, 801 (1986).

\bibitem{Mezard87}
M.\ M\'ezard, G.\ Parisi, and M.\ A.\ Virasoro, 
Spin Glass Theory and Beyond, 
World Scientific, Singapore, 1987.

\bibitem{Nishimori01}
H.\ Nishimori,
Statistical Physics of Spin Glasses and Information Theory, 
Clarendon Press, Oxford 2001.

\bibitem{Viana85}
L.\ Viana, A.J.\ Bray,
Phase diagrams for dilute spin glasses,
J.\ Phys.\ C: Solid State Phys.\ 18, 3037 (1985).


\bibitem{Erdos59}
P.\ Erd\"os and A. R\'enyi, 
On random graphs,
Publicationes Mathematicae 6, 290 (1959).

\bibitem{Barabasi99}
A.-L.\ Barab\'asi, R.\ Albert,
Emergence of scaling in random networks,
Science 286, 509 (1999).

\bibitem{Binder97}
K.\ Binder, D.\ Heermann, 
Monte Carlo Simulation in Statistical Physics, 
Springer-Verlag, Berlin, 1997.

\end{thebibliography}
\end{document}